\documentclass[floats,floatfix,amssymb,prd,aps,twocolumn,nofootinbib,nolongbibliography,reprint,groupedaddress]{revtex4-2}

\usepackage{ulem} % tempor

\usepackage{amssymb,amsmath,verbatim,mathtools,needspace,enumitem,etoolbox,graphicx,physics,microtype,afterpage,xspace,tabularx,lmodern,multirow}
\usepackage{gensymb}
\usepackage[dvipsnames, usenames]{xcolor}
\usepackage[unicode, colorlinks=true, linkcolor=linkcolor, citecolor=linkcolor, filecolor=linkcolor, urlcolor=linkcolor, linktocpage, breaklinks]{hyperref}
\newcommand{\orcid}[1]{\href{https://orcid.org/#1}{\includegraphics[width=10pt]{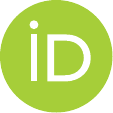}}}
\usepackage[nolist,nohyperlinks]{acronym}
\usepackage{tikz}
\usetikzlibrary{decorations.markings}
\usetikzlibrary{decorations.pathmorphing}

\newcommand{\be}{\begin{equation}}
\newcommand{\ee}{\end{equation}}

\definecolor{linkcolor}{rgb}{0.0,0.3,0.5}
\definecolor{olive}{rgb}{0.4,0.7,0.2}
\newcommand{\Eeffmrg}{\hat{E}_{\mathrm{eff}}^{\mathrm{mrg}}}

\newcommand{\bmrg}{\hat{b}_{\mathrm{mrg}}}

\def\i{{\rm i}}
\def\e{{\rm e}}

\allowdisplaybreaks

\begin{document}

\pagenumbering{arabic}

%------------------------------------------------------------------------------

\title{Ringdown amplitudes of nonspinning eccentric binaries}

\author{Gregorio Carullo \orcid{0000-0001-9090-1862}}
\affiliation{Niels Bohr International Academy, Niels Bohr Institute, Blegdamsvej 17, 2100 Copenhagen, Denmark}
\affiliation{School of Physics and Astronomy and Institute for Gravitational Wave Astronomy, University of Birmingham, Edgbaston, Birmingham, B15 2TT, United Kingdom}

\begin{abstract}

    Closed-form expressions for the ringdown complex amplitudes of nonspinning unequal-mass binaries in arbitrarily eccentric orbits are presented.
    They are built upon 237 numerical simulations contained within the RIT catalog, through the parameterisation introduced in [Phys. Rev. Lett. 132, 101401].
    Global fits for the complex amplitudes, associated to linear quasinormal mode frequencies of the dominant ringdown modes, are obtained in a factorised form immediately applicable to any existing quasi-circular model.
    Similarly to merger amplitudes, ringdown ones increase by more than 50\% compared to the circular case for high impact parameters (medium eccentricities), while strongly suppressed in the low impact parameter (highly eccentric) limit.
    Such reduction can be explained by a transition between an ``orbital-type'' and an ``infall-type'' dynamics.
    The amplitudes (phases) fits accuracy lies around a few percent (deciradians) for the majority of the dataset, comparable to the accuracy of current state-of-the-art quasi-circular ringdown models, and well within current statistical errors of current LIGO-Virgo-Kagra ringdown observations.
    These expressions constitute another building block towards the construction of complete general relativistic inspiral-merger-ringdown semi-analytical templates, and allow to extend numerically-informed spectroscopic analyses beyond the circular limit.
    Such generalisations are key to achieve accurate inference of compact binaries astrophysical properties, and tame astrophysical systematics within observational investigations of strong-field general relativistic dynamics.
    
\end{abstract}

\maketitle

\section{Introduction}

Transient gravitational-wave (GW) signals detected by the LIGO-Virgo-Kagra network~\cite{AdvLIGO, AdvVirgo, KAGRA:2020tym} allow to investigate a variety of scenarios with high relevance for astrophysics, cosmology and fundamental physics~\cite{KAGRA:2021vkt,LIGOScientific:2021sio,KAGRA:2021duu,LIGOScientific:2021aug}.
Such observations have already achieved an impressive number of unexpected discoveries, and hold the promise to revolutionise again each of the above fields in the near future.
Delivering such promise requires unbiased inference of the observed signals, which in turn also rests upon accurate waveform models of black hole (BH) binary mergers including all relevant physical effects.
When developing such models, initial focus was placed in describing the dominant quadrupolar mode for BH progenitors with zero intrinsic angular momentum (also known as ``spin''), then including spin degrees of freedom aligned to the orbital angular momentum.
Subsequently, the target shifted towards the robust modeling of higher angular harmonics (also known as ``higher modes''), and spin degrees of freedom perpendicular to the orbital angular momentum.
The latter case induces a precession of the orbital plane, hence is referred to as the ``spin-precessing case''.
State-of-the-art BH binary models~\cite{Varma:2018mmi,Pratten:2020ceb,Estelles:2021gvs,Gamba:2021ydi,Thompson:2023ase,Ramos-Buades:2023ehm} nowadays include both higher modes and precessional degrees of freedom, to various degrees of approximation.
All of these models are constructed under the assumption that the binary evolves in a series of circular orbits until the plunging stage.
For pointers to the literature of the large body of work that underwent in the construction of such high-accuracy models, see e.g. Refs.~\cite{LIGOScientific:2020stg,LIGOScientific:2020zkf}.

Recently, there has been a surge of interest in enhancing these templates by relaxing the circularity assumption and allowing for generic noncircular orbits.
This is an important extension of current models, since binary systems forming dynamically or evolving in dense environment might enter the latest inspiral cycles with non-zero eccentricity~\cite{Mandel:2018hfr,Mapelli:2021taw}.
Including eccentric contributions would be relevant not only to infer the relative importance of astrophysical formation channels, but also to avoid biases when observationally investigating the general relativistic two-body dynamics, or in searches of new physics.
For a recent review of the relevant literature, see the introduction of Ref.~\cite{Carullo:2023kvj}.
A noteworthy and recent addition to the above review is Ref.~\cite{Gamba:2024cvy}, putting forward an effective-one-body noncircular and precessing waveform model, comparing it against currently publicly available (although quite restricted) set of numerical simulations.
This marks a significant milestone in the construction of robust and complete waveform templates from two-body systems in general relativity.
See their introduction for additional references on the topic.
Despite such remarkable progress in the modeling of the inspiral stage of the signal, all current noncircular closed-form models\,\footnote{A numerical surrogate model valid for small eccentricity was presented in Ref.~\cite{Islam:2021mha}.} rely on the assumption that by the time the binary merges, the system has rapidly circularised due to the emission of GWs~\cite{Sperhake:2007gu,Hinder:2007qu}.
In practise, this means that the eccentric inspiral is ``smoothly stiched'' to a numerically-informed quasi-circular merger-ringdown template, hence lacking noncircular effects.
Albeit such assumption is certainly robust for low initial eccentricity, this is not the case for highly noncircular binaries~\cite{Carullo:2023kvj}.
Additionally, recent analyses of GW data in Refs.~\cite{Romero-Shaw:2020thy,Gamba:2021gap,Gupte:2024jfe}, have pointed to the presence of noncircular degrees of freedom for massive binary systems when employing closed-form templates\,\footnote{For comparisons with numerical simulations, see Refs.~\cite{Gayathri:2020coq,CalderonBustillo:2020xms}.}.
Earlier analyses of bounded eccentric BH binaries were carried in  Refs.~\cite{Bonino:2022hkj,Iglesias:2022xfc}, reporting no evidence for eccentricity in the subset of events analysed.
However, all these analyses suffer from the aforementioned limitation of including a quasi-circular merger-ringdown component.
Since the signal-to-noise ratio of such massive systems is dominated by the merger and ringdown stages, it is important to corroborate the above assumption, and urgently construct complete noncircular models.
The focus on high-mass systems is also justified from an astrophysical perspective, since hierarchial mergers (leaving behind massive remnants) are expected to be a key component of formation channels leading to noncircular binaries~\cite{Gerosa:2021mno}.

In Ref.~\cite{Carullo:2023kvj}, a first step towards this direction was undertaken, constructing closed-form expression for the remnant (final mass and spin) and merger (amplitude and frequency at the waveform peak) quantities for aligned-spin binaries with comparable masses in generic noncircular orbits (i.e. for both bounded and initially unbounded mergers).
These fits constitute one of the building blocks of generic merger-ringdown models, and were made possible by a novel parameterisation which shifted the focus from orbit-based quantities, to dynamics-based ones.

Here, another step in this direction is made by leveraging the same parameterisation to construct closed-form expressions of ringdown amplitudes and phases associated to the linear quasinormal-mode (QNM) frequencies of dominant multipoles, for nonspinning binaries of moderate mass ratios, and arbitrary initial eccentricities.

The paper is structured as follows.
Conventions are set in Sec.~\ref{sec:conventions}, while the employed Kerr ringdown model is described in Sec.~\ref{sec:model}. 
Extraction of the ringdown complex amplitudes, including a description of the numerical dataset are presented in Sec.~\ref{sec:amplitude_extraction}.
Global fits as a function of noncircular parameters are constructed in Sec.~\ref{sec::global_fit}, where a physical interpretation of the results is also discussed. 
Finally, Sec.~\ref{sec:conclusions} contains a discussion of future developments and concluding remarks.

\section{Conventions}\label{sec:conventions}

Geometric units $c=G=1$ are used throughout the manuscript.
A spin-weighted spherical harmonics $_{-2}Y_{\ell m}(\theta,\varphi)$ decomposition of the GW strain at infinity is considered

\be
    h_+-{\rm i}h_\times =  \sum_{\ell,m} h_{\ell m}(t) {}_{-2}Y_{\ell m}(\theta,\varphi) \,.
\ee

Each spherical mode $h_{\ell m}(t)$ can also be conveniently split in amplitude and phase as

\be
    h_{\ell m}(t) = A_{\ell m}(t)\,\e^{i \phi_{\ell m}(t)},
\ee

and the instantaneous GW frequency is defined as:

\be
    f_{\ell m}(t) \equiv \dot{\phi}_{\ell m}(t) / 2\pi  \, .
\ee

Individual initial horizon (Christodoulou) masses of the two BHs are denoted as $m_{1,2}$, with $M$ = $m_1 + m_2$, the mass ratio $q \equiv m_1 /m_2 \geq 1$, and the symmetric mass ratio $\nu = m_1 m_2 / M^2$.
The time-axis is constructed by setting $t=0$ at the peak of $h_{22}$, and quoted in units of the initial Arnowitt-Deser-Misner mass of the spacetime, set to unity, $M=1$.

\section{Waveform model}\label{sec:model}

For a given harmonic, a template composed of a superposition of damped sinusoids and a tail term is constructed, and interfaced through the \texttt{pyRing} package~\cite{pyRing}.
Such general ringdown template reads

\begin{eqnarray}\label{eq:template}
    h^{\rm Kerr}_{\ell  m} (t) &=& \sum_{\ell^\prime=2}^{\infty} \sum_{n=0}^{\infty} \left[ A^{+}_{\ell^\prime m n} \, e^{i ( \omega^{+}_{\ell^\prime m n}(t-t_{\rm ref})+\phi^{+}_{\ell^\prime m n} ) }  \right. \,\, \nonumber\\
    &+& \left. A^{-}_{\ell^\prime m n} \, e^{i(\omega^{-}_{\ell^\prime m n}(t-t_{\rm ref})+\phi^{-}_{\ell^\prime m n})} \right] \theta(t-t^{\rm ring}_{\rm start})  \nonumber\\
    &+& A^T_{\ell m} \, e^{i \phi^T_{\ell m}} \, (t-t_{ref})^{p^T_{\ell m}}  \,\, \theta(t-t^{\rm tail}_{\rm start}) \, ,
\end{eqnarray}
with $\ell>2$, $m\geq0$, where $\omega^{\pm}_{\ell m n}$ correspond to Kerr complex QNM frequencies fixed by the final BH mass and spin, interfaced using the \texttt{qnm} package~\cite{Stein:2019mop}.
The constant quantities $(A^{\pm}_{\ell m n}, \phi^{\pm}_{\ell m n})$ are a priori unknown for comparable-mass mergers, and their extraction constitutes the main goal of this study.
Negative-$m$ modes are constructed from positive-$m$ ones through reflection symmetry $h_{\ell,-m} = (-1)^{\ell} h^{*}_{\ell,m}$, implied by the symmetries of the Kerr metric and non-precessing initial data~\cite{Blanchet:2013haa, London:2018gaq}.
Without loss of generality, the amplitudes are imposed to be positive-defined.

The equation above contains several contributions. 
First, an a priori infinite number of modes with different $\ell'$, but same $m$, will contribute to each $\ell$ harmonic, due to spherical-spheroidal mixing~\cite{Berti:2005gp,Buonanno:2006ui,Kelly:2012nd,Berti:2014fga,London:2018nxs}.
Among the highest excited modes, this mixing is particularly predominant for the $(3,2,0)$ mode, and has only a small impact on the other modes considered in this work.
Second, two family of modes will contribute to each harmonic, physically corresponding to perturbations that respectively co-rotate (``+'') or counter-rotate (``-'') with respect to the BH intrinsic angular momentum $a_f$, see e.g. Refs.~\cite{Berti:2005ys,Lim:2019xrb,Li:2021wgz}.
Co-rotating modes have $\Re{\omega^{+}_{\ell  m  n}} > 0$, while  ``counter-rotating'' modes have $\Re{\omega^{-}_{\ell  m n}} < 0$.
For binaries, the latter are appreciably excited to non-negligible levels when the perturbation of the remnant BH (controlled by the binary orbital angular momentum $L$, here taken to be along the z axis, hence dropping the vector notation) is counter-aligned with respect to the remnant BH angular momentum $a_f$, or for low remnant spins.
For binaries in quasi-circular planar orbits, these configurations arise for high intrinsic binary spins $(a_1, a_2)$ pointing in the opposite direction of $L$, or high mass ratios.
For the nonspinning and noncircular dataset under consideration, they are important only for extremely high eccentricities (beyond $e_0 \simeq 0.9$) that give rise to a low-spin remnant.
Since these values of eccentricity are not included in our dataset for data quality reasons (see below), counter-rotating contributions will not be considered, and the superscript "+" will be suppressed when considering co-rotating modes.
It was explicitly verified that adding these contributions does not appreciably alter the results presented in this work.
In principle, one should also consider a tail term, for which the values of $(A^T_{\ell m} \, \phi^T_{\ell m}, p^T_{\ell m})$ would also need to be extracted.
Typically, these quantities have been taken as constant~\cite{Carullo:2023tff, Baibhav:2023clw}.
However, recent studies~\cite{Albanesi:2023bgi, DeAmicis:2024not, Islam:2024vro} have shown that in source-driven relaxations, as is the case for binaries, a regime in which a single pure power-law dominates is reached only at very late times (typically $10^5$).
Hence, it would be necessary to promote these quantities to time-dependent functions, or consider a superposition of a very large number of power-laws in order to obtain an accurate fit~\cite{DeAmicis:2024not}.
Albeit the above considerations are important for future high-accuracy studies on tail extraction, they will not enter the fit construction below, since tail terms can have an impact only for $e_0 \geq 0.9$ (within the available dataset accuracy levels), hence in what follows the tail term is neglected as well.
Due to simulations inaccuracies, for certain simulations the strain does not converge towards zero, but rather to a small constant value.
This is a typical systematic error present in numerical simulations (see e.g.~\cite{Cheung:2023vki}), and is accounted for by simultaneously fitting for a constant term.
Hence, in practise, the tail term is used to capture this constant and remove its influence on the QNM amplitudes extraction, fixing in the template above $p^T_{\ell m n} = 0$, hence subtracting a constant complex value equal to $A^T_{\ell m} \, e^{i \phi^T_{\ell m}}$.

Finally, $t_{\rm ref}$ represents the time at which complex amplitude parameters are defined.
Given the dependence of $\omega_{\ell m n}$ on the BH spin, different choices of $t_{\rm ref}$ rescale the amplitude factors of different binary configurations in a non-trivial way.
In Ref.~\cite{London:2018gaq}, it was found that setting reference time $t_{\rm ref}$ to the maximum of $|h_{22}|$, $t_{\rm ref} = t_{\rm peak}$, allows to simplify the complex amplitude dependence on binary parameters.
Here, the same logic is followed, while accounting for the fact that in highly eccentric scenarios the peak of the waveform does not in general provide a good estimate of the merger time.
This is because for intermediate eccentricity values, the last periastron before merger can display a larger amplitude compared to the merger itself, as displayed e.g. in Fig.~\ref{fig:Amp_ecc}.
See also Figs.~1,~6 of Ref.~\cite{Nagar:2020xsk} or Supplementary Fig.~3 of Ref.~\cite{Gamba:2021gap}, diplaying conceptually identical phenomenologies.
Hence, the definition $t_{\rm ref} = t_{\rm mrg}$ is used, where $t_{\rm mrg}$ is the last peak of $A_{22}$ immediately before a ringdown regime begins, an appropriate estimate of the merger time.
For an extended discussion of this point, see Ref.~\cite{Carullo:2023kvj}.
This choice allows to automatically obtain peak-rescaled amplitude values when varying the fit start time.
The above $t_{\rm ref}$ parameter should not be confused with $t_{\rm start}$\,\footnote{Differently from QNM, which are excited during light-ring crossing, tails are mostly excited during apastra~\cite{DeAmicis:2024not}, hence will have in general a different starting time. Since tail terms will not be considered here, this point will not enter the fitting procedure below, hence for ease of notation $t_{\rm start}=t^{\rm ring}_{\rm start}$}, setting the validity regime of a model built through a superposition of QNMs with constant mass, spin and complex amplitude.
Earlier times are contaminated by transients~\cite{Berti:2006wq,Lagos:2022otp,Albanesi:2023bgi}, non-linearities~\cite{Gleiser:1996yc,London:2014cma,Sberna:2021eui,Cheung:2022rbm,Mitman:2022qdl, Lagos:2022otp, Baibhav:2023clw,Cheung:2022rbm,Bucciotti:2023ets, Perrone:2023jzq, Redondo-Yuste:2023seq, Ma:2024qcv,Bucciotti:2024zyp,Bourg:2024jme,Zhu:2024rej} and mass-spin variations~\cite{Sberna:2021eui,Redondo-Yuste:2023ipg,Zhu:2024dyl,May:2024rrg,Capuano:2024qhv}, for which no first-principles model exists at present, with current templates relying on phenomenological constructions~\cite{Damour:2014yha,Bohe:2016gbl,Cotesta:2018fcv,Estelles:2020osj,Estelles:2021gvs,Estelles:2020twz,DelPozzo:2016kmd, Nagar:2018zoe, Nagar:2019wds, Nagar:2020pcj}.
Hence, the scope of the present investigation will be limited to construct a model for the late-time \textit{constant} amplitudes, possessing a robust theoretical base and a clean physical interpretation.

\section{Complex amplitudes extraction}\label{sec:amplitude_extraction}

\subsection{Numerical dataset}

The considered dataset is restricted to binaries with nonspinning progenitors and contains 237 noncircular bounded simulations in arbitrarily eccentric orbits, available in the fourth public release of the RIT catalog of waveforms extrapolated at future null infinity~\cite{Healy:2022wdn}.
The same nonspinning dataset of Ref.~\cite{Carullo:2023kvj} is employed, with the mass ratio ranging in $q=[1,3]$.
Details of the simulations and of selection criterion used, based on the quality of the numerical data (quantified by balance laws), can be found in the latter reference.
Particularly, data quality considerations limit the available dataset to binaries with $e_0 \leq 0.9$.

\begin{figure*}[thbp]
\centering
    \includegraphics[scale=1, width=\textwidth]{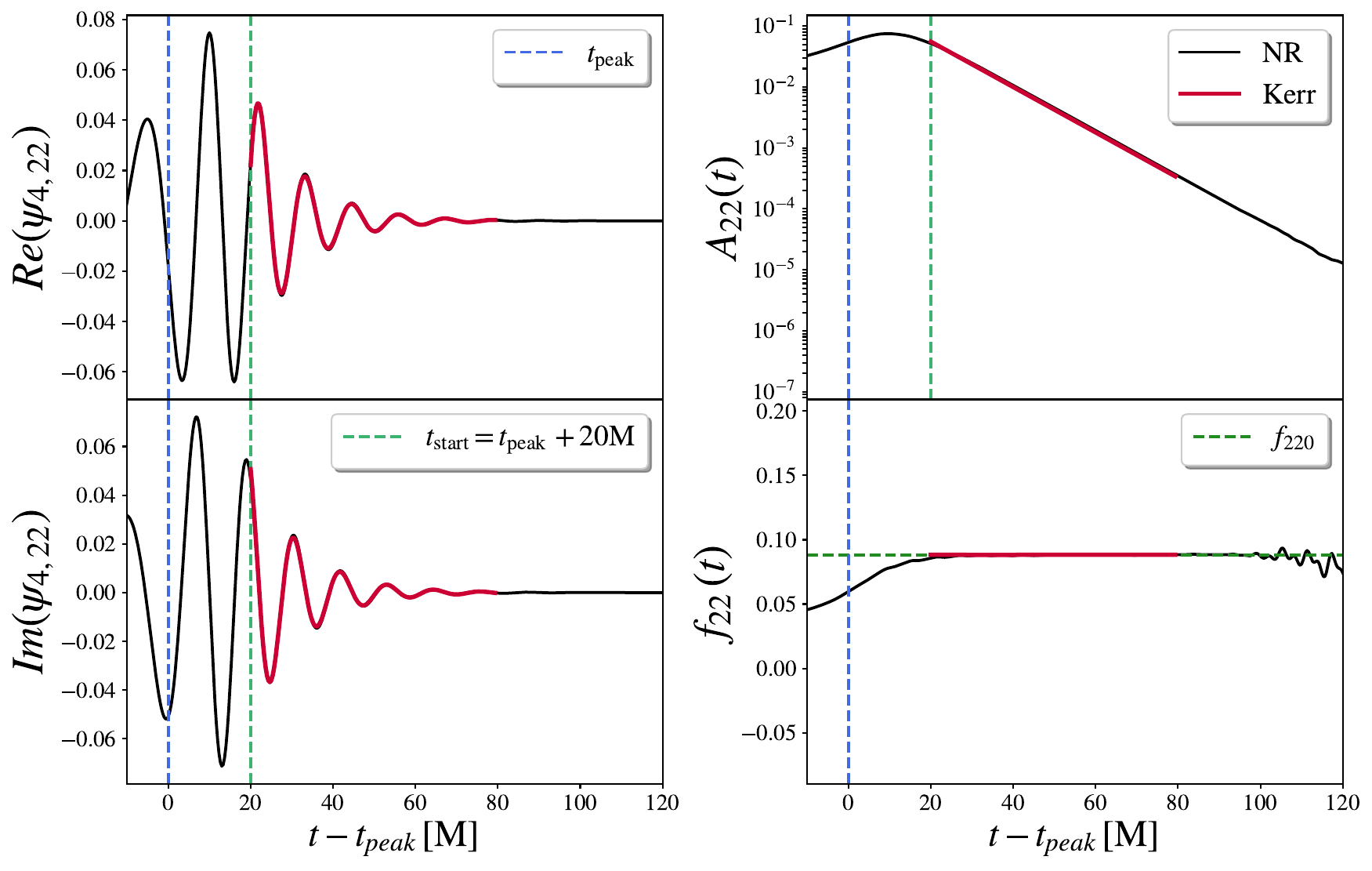}       
   \caption{Quadrupolar mode of the $\psi_4$ simulation output for the illustrative case \texttt{RIT:1334}, with $e_0 \simeq 0.5$ and $q \simeq 1.1$ (black).
   The red line represents an illustrative fit using only the fundamental ringdown mode with free amplitude and phase, fitting the numerical data only after $t_{\rm start}$.
   The peak time $t_{\rm peak}$ refers to the peak of $h_{22}$.
   The green horizontal line represents the value of $f_{220} \equiv \omega_{\ell m n}/2\pi$.
   }
   \label{fig:single_sim_waveform}
\end{figure*}

Results presented below are constructed by directly fitting the strain data contained in the catalogue release, obtained from double integration of the Weyl scalar $\psi_4$, constituting the direct simulation output.
To test against systematics induced by the strain reconstruction procedure~\cite{Hopper:2022rwo,Andrade:2023trh,Albanesi:2024xus}, the fits of the complex amplitudes were repeated on $\psi_4$, converted to strain amplitudes by appropriately rescaling using the QNM frequencies, as done in e.g. Refs.~\cite{London:2014cma, Redondo-Yuste:2023seq}.
Indistinguishable results in the amplitudes of the modes considered were found, confirming the robustness of the employed extraction procedure.
An illustrative case of the typical quality of a the actual simulation output with medium eccentricity ($e_0 = 0.51$) and close to equal-mass $q=1.11$, \texttt{RIT:1334}, is displayed in Fig.~\ref{fig:single_sim_waveform}.

\subsection{Extraction procedure}

From the above model, two free parameters per each harmonic mode $(A_{\ell m 0}, \phi_{\ell m 0})$ need to be extracted.
As mentioned above, when considering higher modes the nuisance parameters $A^T_{\ell m} \, \phi^T_{\ell m}$ are also added, accounting for the late-time constant observed in certain simulations due to numerical inaccuracies.

The fit is performed using a Bayesian method.
The algorithm, implemented in the \texttt{bayRing} package~\cite{bayRing}, is described in detail in Ref.~\cite{Redondo-Yuste:2023seq} and relies on the \texttt{cpnest} sampling algorithm~\cite{cpnest}.
Given the simple inference problem at hand, 32 live points and maximum Markov Chain Monte Carlo steps, with a single chain, are used.
This values are increased to 64 when considering mode-mixing contributions, which double the number of free parameters.
Such sampler settings were tested to be sufficient to obtain an unbiased extraction, as the results did not change when increasing sampler settings.
Wide parameters ranges are explored, with priors uniform in: $[0, 2\pi]$ on phase parameters, $[-20, 5]$ on $ln \, A^{\pm}$ and $ln \, A^{T}$, for all the modes.

Since RIT waveforms come with a single resolution and extraction radius, a constant gaussian error estimate on the GW strain is assumed.
A conservative estimate of the a noise floor is selected, $\sigma=10^{-4}$, which sets the precision threshold required in the fit, and data beyond $t_{\rm end} = t_{\rm peak} + 80 \, M$ are excise from the fit.
The latter choice only removes a very small portion of signal power for lower eccentricities, while preventing contamination from simulation artifacts at late times for all cases considered, including the ones with lower data quality.
If multiple resolutions and extraction radii will be available in the future, the above error could be instead promoted to a time-dependent quantity, and sub-dominant modes could have been extracted, as demonstrated in Ref.~\cite{Redondo-Yuste:2023seq}.
Lacking such improved estimates, the error is set to the smallest value possible, while avoiding overfitting features that could be induced by the numerical error.

The model includes only the dominant $(\ell, m, 0)$ fundamental mode for each of the $(\ell, m) = [(2,2), (3,3), (2,1), (4,4)]$ spherical harmonics considered.
For the $(\ell, m) = (3,2)$ harmonic, the dominant mode-mixing contribution from the $(\ell,m,n) = (2,2,0)$ mode was also included, in addition to the $(3,2,0)$.
Attempts to construct a robust global fit the overtones and counter-rotating contributions, expected to constitute the next dominant component, failed due to the limited available data quality.
Similarly, a robust fit of the $(\ell,m,n) = (2,0,0)$ mode could not be obtained.
This is expected since we do not include eccentricity values close to unity, which is the regime where this mode is expected to be strongly excited, and since memory contributions (dominant for such mode~\cite{Mitman:2024uss}) are not accurately resolved in the dataset under consideration.

\subsection{Single binaries results: start time variation}

Overfitting is assessed by repeating the fits at different starting times, keeping the end time fixed, and extracting the amplitudes values when they are approximately constant, as predicted by perturbation theory once the initial transient has decayed~\cite{London:2014cma, Lagos:2022otp, Baibhav:2023clw}.
The start time of the fit $t_{\rm start}$ is thus varied in the interval $[0,40] \, M$ in steps of $2 \, M$.

\begin{figure*}[thbp]
\centering
    \includegraphics[scale=1, width=0.48\textwidth]{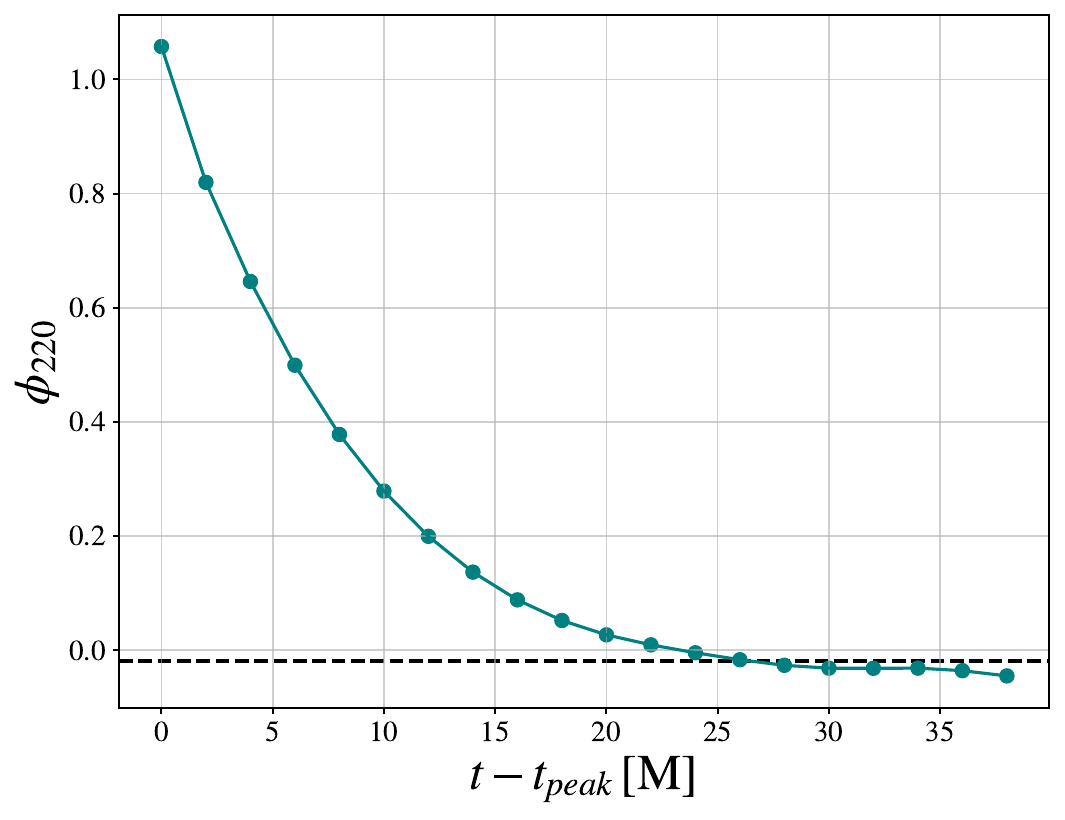}       
    \includegraphics[scale=1, width=0.48\textwidth]{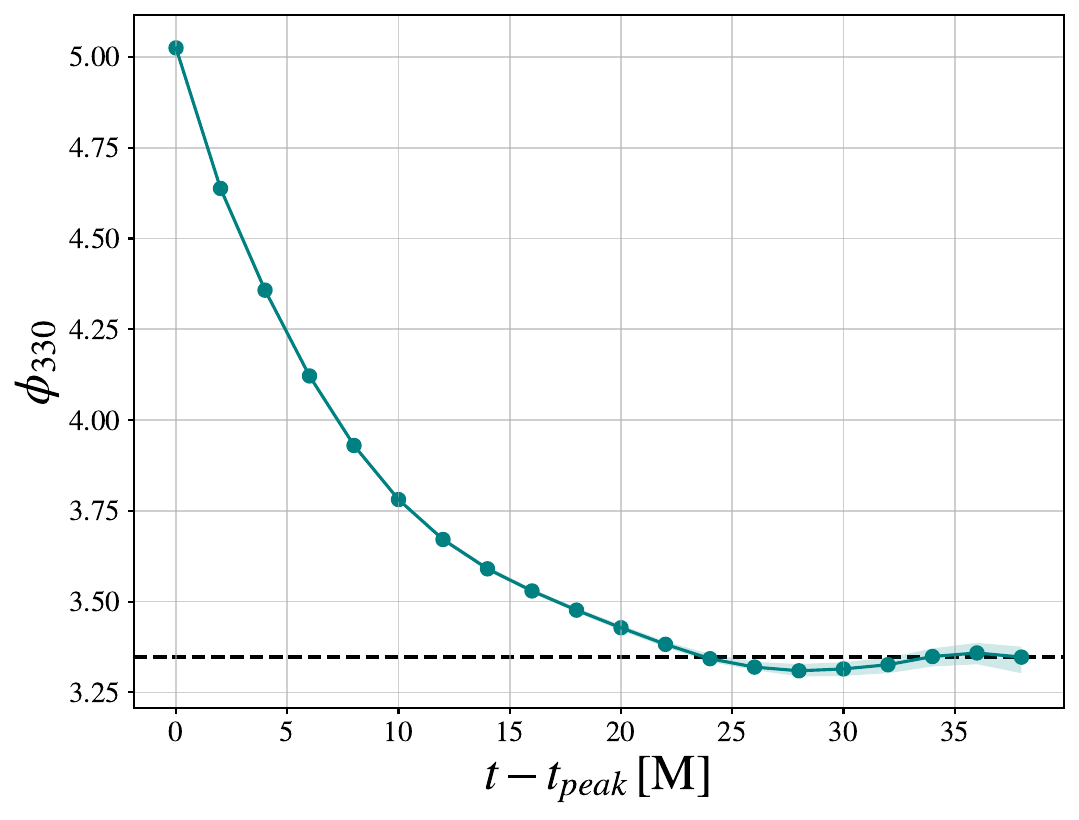}    
    \includegraphics[scale=1, width=0.48\textwidth]{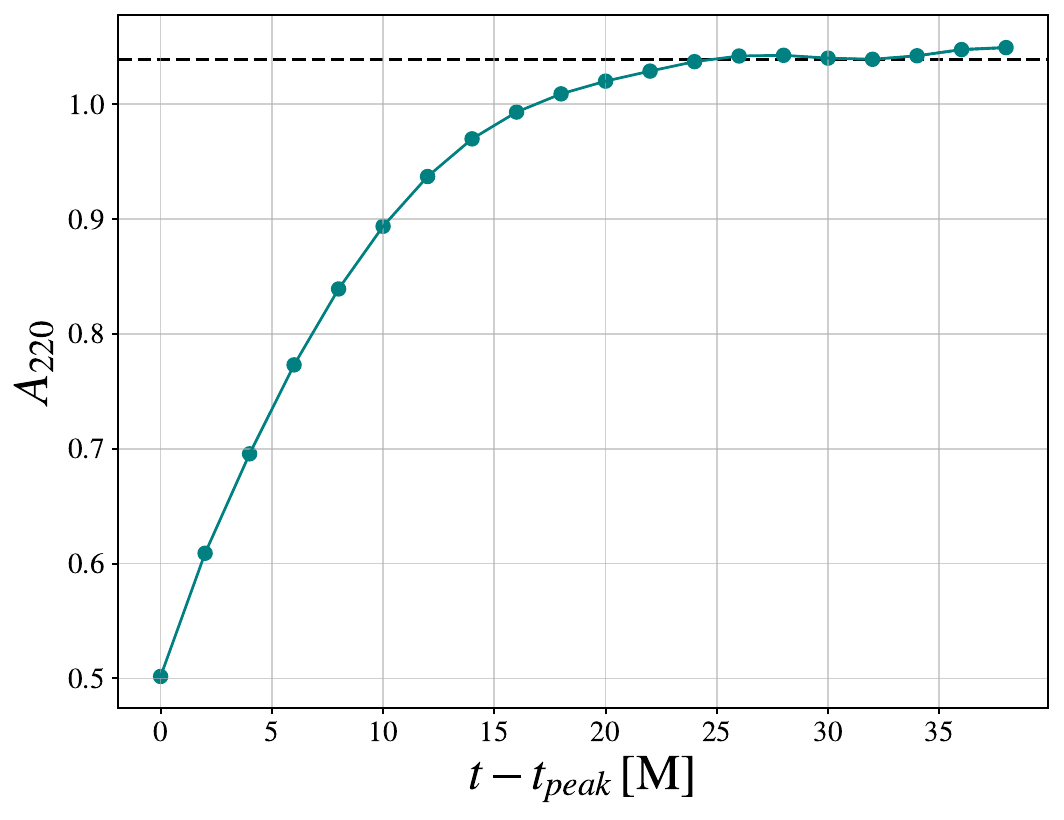}   
    \includegraphics[scale=1, width=0.48\textwidth]{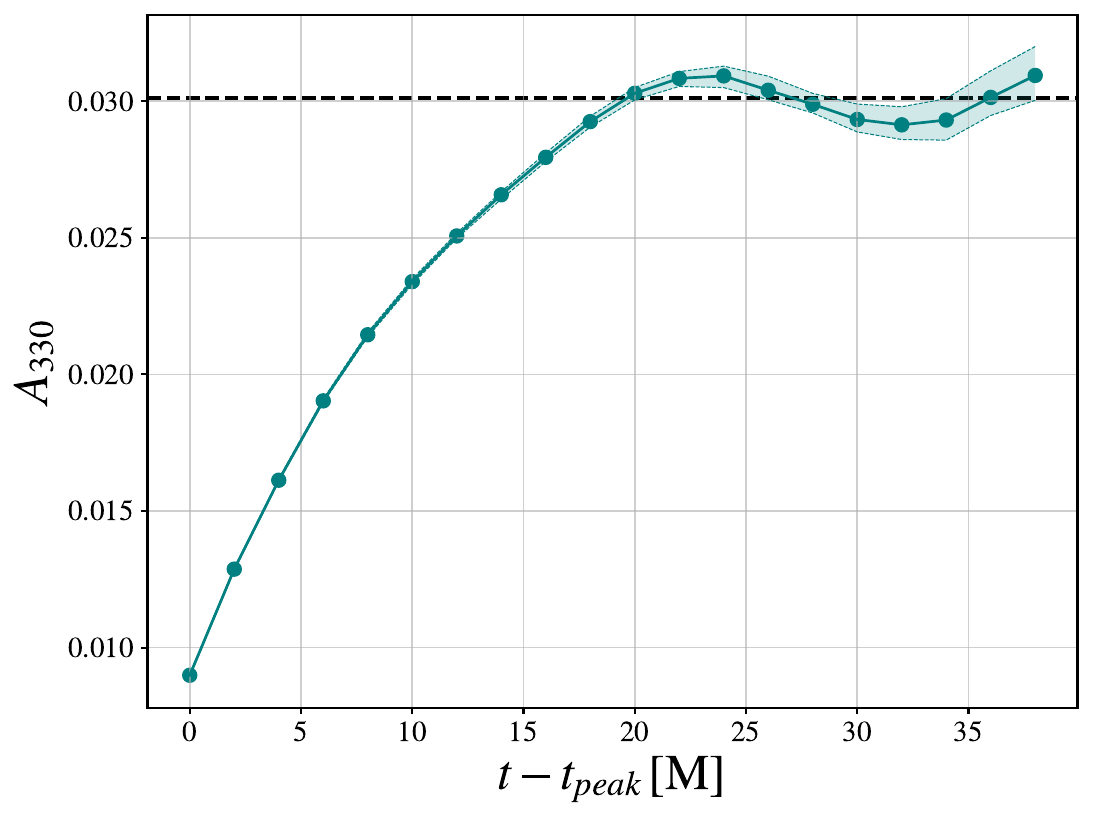}
   \caption{
   Amplitude and phase parameters extraction at different starting times, for the illustrative case \texttt{RIT:1334}, with $e_0 \simeq 0.5$ and $q \simeq 1.1$.
   The solid teal line represents the median of the posterior at each start time, emphasized by filled dots.
   Thin dashed teal lines represent the 90\% credible levels of each posterior (shaded).
   The dashed horizontal black line represent the constant fit starting from $20\,M$.
   Errors in the $(2,2,0)$ mode are too small to be visible.
   }
   \label{fig:single_sim}
\end{figure*}

The results of the fit at multiple starting times is presented in Fig.~\ref{fig:single_sim} for the illustrative case \texttt{RIT:1334}.
As an example, both the dominant quadrupolar $(2,2,0)$ mode and the dominant octupolar mode $(3,3,0)$ are shown.
Results for other modes are qualitatively identical.
As expected, the amplitude initially grows, then saturates to an approximately constant value.
The growth is expected to be driven by transient effects, with contaminations from non-linearities, additional fast-decaying modes, non-mode contributions~\cite{Zhu:2023mzv}, and mass-spin variation (see the discussion in Sec.~\ref{sec:model}).
All of these effects seem to disappear after $\sim 20 \, M$, when the single mode amplitude stabilises towards a constant.
This is expected at least when considering mass-spin changes, which become negligible on this timescale~\cite{Buonanno:2006ui,Berti:2007fi,May:2024rrg}, and fast-decaying modes~\cite{Baibhav:2023clw}.
The phase shows a similar behaviour.
Note that the phases values at different starting times have been ``unwrapped'', namely certain values have been shifted using $\phi \rightarrow \phi + 2\pi$, in order to enforce continuity between all adjacent points.
This operation is always allowed given the phases periodicity.
This step is required since by fit construction the phases obtained will always be within $[0, 2\pi]$, and a phase value close to e.g. $2\pi$ at a given start time might jump to zero at the next starting time, giving rise to a discontinuity.
When constructing closed-form models, is preferable to avoid such discontinuity.
After this unwrapping step, the resulting phase values as a function of $t_{\rm start}$ will be continuous, but now not necessarily within $[0, 2\pi]$ anymore.
Subsequent unwrapping stages will be applied in the construction of the global fit, described below.

To extract an estimate of the physical QNM amplitude, the medians of the amplitude posteriors obtained at the different start times values $t_{\rm start} \in [20,40] \, M$ are fit with a constant.
The lower limit is consistent with the choices of previous studies~\cite{London:2018gaq}.
The upper limit serves to exclude contributions from times where the numerical noise can start to dominate for the most challenging extractions.
The choice of this time window serves to:
i) moderate biases due to the small residual evolution of the waveform frequency, which has still not sufficiently converged to the asymptotic QNM value, hence might still slowly evolve even after $20\,M$;
ii) average over possible contaminations due to sub-dominant modes, which could not be smoothly resolved across the dataset; 
iii) average over variations induced by numerical inaccuracies.
Oscillations due to these contamination factors, similarly to what displayed in the right panels of Fig.~\ref{fig:single_sim},  are especially present for the high-eccentricity cases.

In future work, the latest step should be generalised to a Bayesian fit, obtaining a posterior distribution on the amplitude value, which would propagate in a posterior distribution over the global fit constructed in the next section, similar to what was done e.g. in Ref.~\cite{Carullo:2021oxn} for the QNM spectrum, or at a higher level in waveform models constructions in e.g. Refs.~\cite{Williams:2019vub,Breschi:2022ens,Pompili:2023tna}.
This procedure would help in mitigating the systematic error due to the constant amplitude extraction, although for the dominant modes it might be made unnecessary by improvements in the data quality of the simulation dataset.

\begin{figure*}[thbp]
\centering
    \includegraphics[scale=1, width=0.45\textwidth]{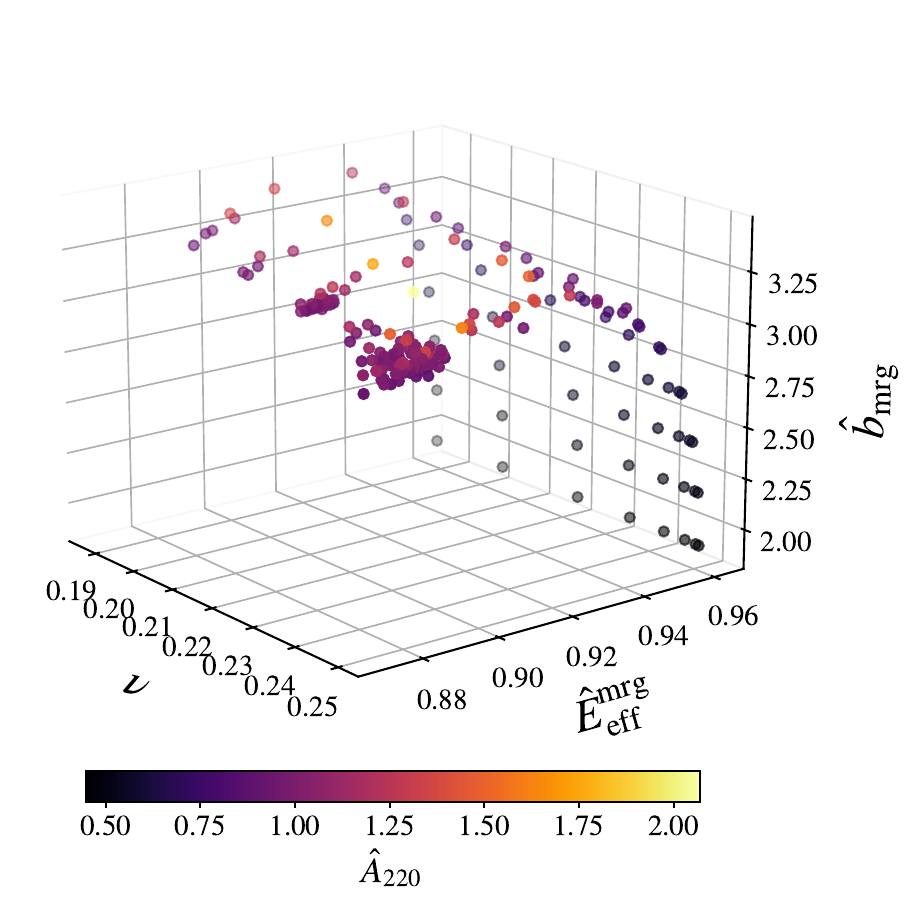}   
    \includegraphics[scale=1, width=0.45\textwidth]{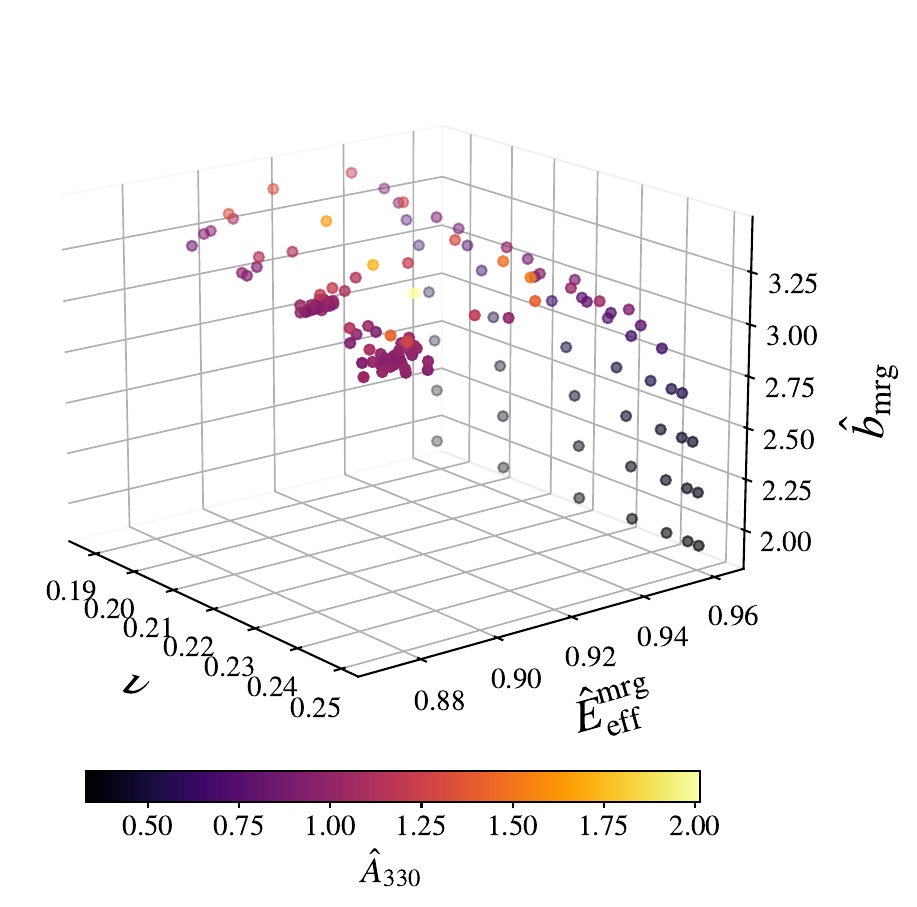}
    \includegraphics[scale=1, width=0.45\textwidth]{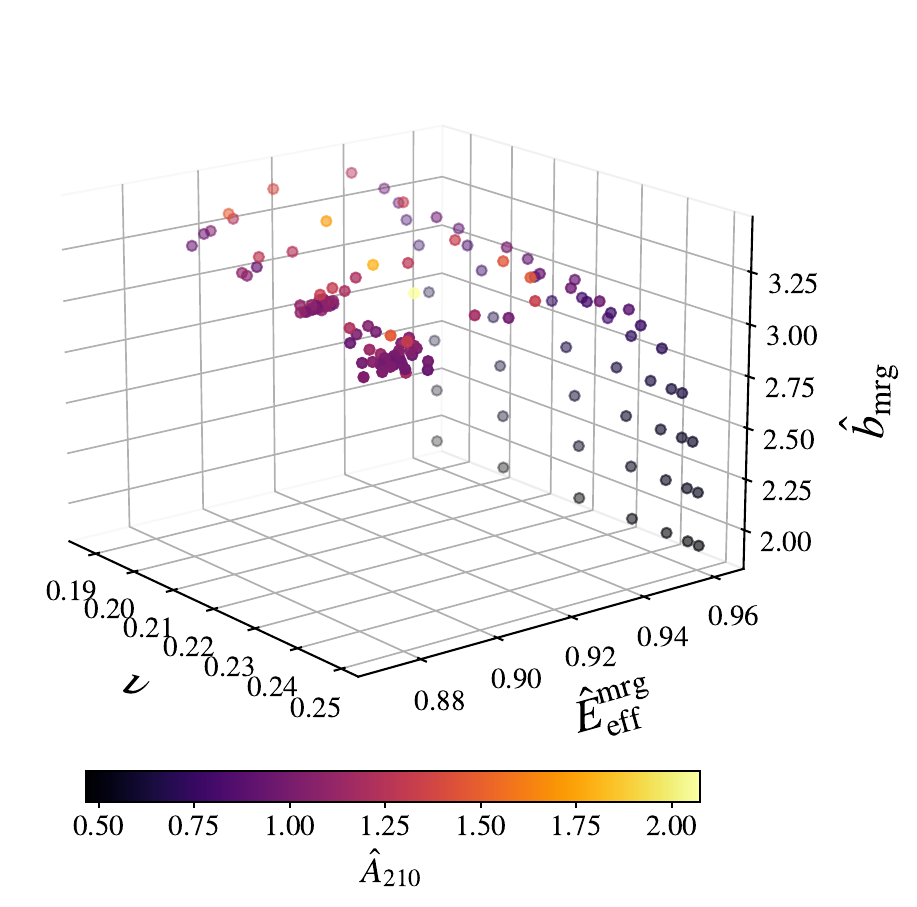}   
    \includegraphics[scale=1, width=0.45\textwidth]{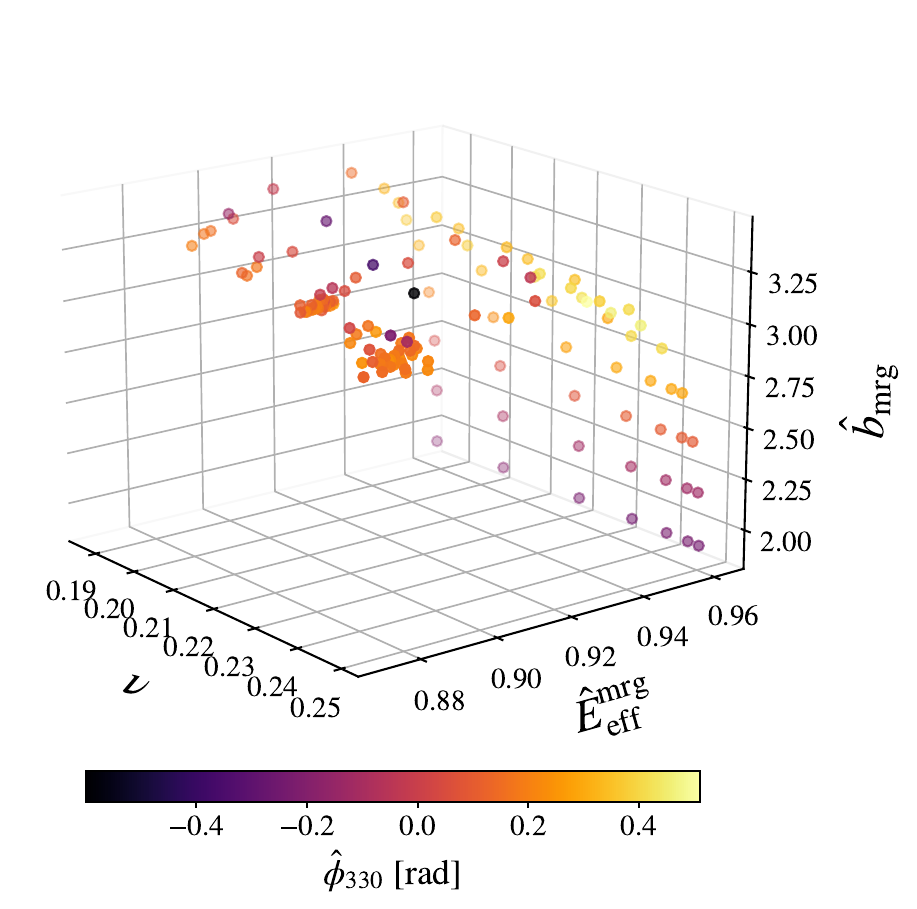}
    \includegraphics[scale=1, width=0.45\textwidth]{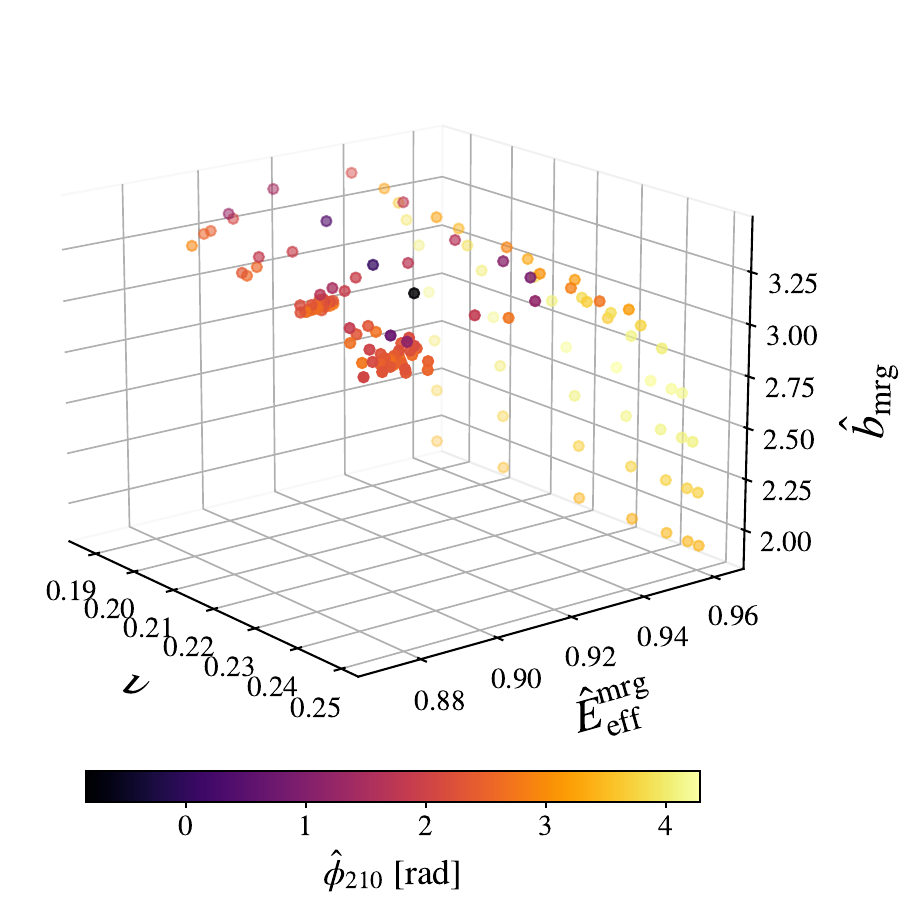}
   \caption{
   Parameter space of three dimensional fits, in terms of two noncircular parameter and mass ratio.
   }
   \label{fig:3D_dataset}
\end{figure*}

\section{Global fit}\label{sec::global_fit}

In this section, the procedure to construct a ``global fit'' in terms of the progenitors binary parameters for the complex amplitudes extracted above, is presented.
First, the gauge-invariant parameterisation introduced in Ref.~\cite{Carullo:2023kvj}, used to represent the noncircular initial conditions, is summarised.
Then, the fit details and the results obtained are reported.

\subsection{Noncircular parameterisation}\label{subsec:par}

Extraction of complex amplitudes was already achieved in Ref.~\cite{Forteza:2022tgq}, in the context of searches of new physics.
However, the amplitude dependence on eccentricity was found to be non-bijective, preventing a global fit construction for the eccentric sector.
This obstacle, which is common to constructing closed-form expression for the merger and remnant quantities, has been overcome in Ref.~\cite{Carullo:2023kvj}.
There, a new gauge-invariant parameterisation, shifting the perspective from orbital-based parameterisations to dynamics-based ones was introduced.
Specifically, starting from the ADM energy and angular momentum calculated at the beginning of the simulation $(E^{\rm ADM}_0, J_0^{\rm ADM})$ and taking into account the GW losses, it is possible to compute the values of the mass-rescaled (adimensional) effective energy and angular momentum at merger $(\Eeffmrg, \, j_{\mathrm{mrg}})$.
These quantities are defined at $t_{\mathrm{mrg}}$, introduced in Sec.~\ref{sec:model}.
A convenient combination of these variables in the form of an effective ``impact parameter'' at merger, $\bmrg$, can also be constructed.
The latter parameter was previously considered for bound orbits in the test-mass limit within Ref.~\cite{Albanesi:2023bgi}, and in Ref.~\cite{Carullo:2023kvj} was found to yield unexpectedly simple and accurate expresssions of the merger properties and of the remnant BH parameters $(M_f, \, a_f)$, when including appropriate comparable mass corrections.
Moreover, although in principle two quantities would be needed to represent the phase space of noncircular binaries, the single quantity $\bmrg$ seems to capture very accurately the noncircular dependence of the merger-remnant parameters, yielding bijective expressions (unlike e.g. eccentricity-like parameters, even when using gauge-invariant definitions).
This property, was dubbed a form of ``quasi-universality'' in Ref.~\cite{Carullo:2023kvj}, to convey the fact that two systems with different initial energy and angular momentum, but similar impact parameter at merger, will give rise to closely to indistinguishable merger-remnant properties.
This feature has yet to find a formal derivation. 
However, in an EOB picture~\cite{Buonanno:2000ef} or for test-mass systems~\cite{Ori:2000zn}, it is known that the properties of the plunge-merger dynamics are determined by the radius of the last stable orbit.
It seems then plausible to conjecture that such radius can be determined by the impact parameter at merger.
Hence, an extension of the analyses of Refs.~\cite{Buonanno:2000ef,Ori:2000zn} to noncircular orbits, or a similar extension of the recent results of Ref.~\cite{Kuchler:2024esj}, seems the most promising avenue to explain such quasi-universal behaviour.

Beyond quasi-universal parameterisations in terms of $\bmrg$, the merger amplitude was found to be well-described in terms of $(\bmrg, \Eeffmrg)$.
Since this study also deals with amplitude parameters, the same parameterisation will be adopted here.
The fits have also been performed with alternative variable combinations (including e.g. the angular momentum) or in terms of $(M_f, a_f)$ without finding any significant simplification.
Similarly, rescaling by the merger amplitude did not yield appreciable simplifications.
Hence, in what follows, the set of variables $(\nu, \, \bmrg, \, \Eeffmrg)$ will be used to construct the global fit.
The reader should refer to Ref.~\cite{Carullo:2023kvj} for details on the fluxes computations entering the above definitions, and for additional motivation behind the parameterisation summarised above.

\begin{figure*}[thbp]
\centering
    \includegraphics[scale=1, width=0.7\textwidth]{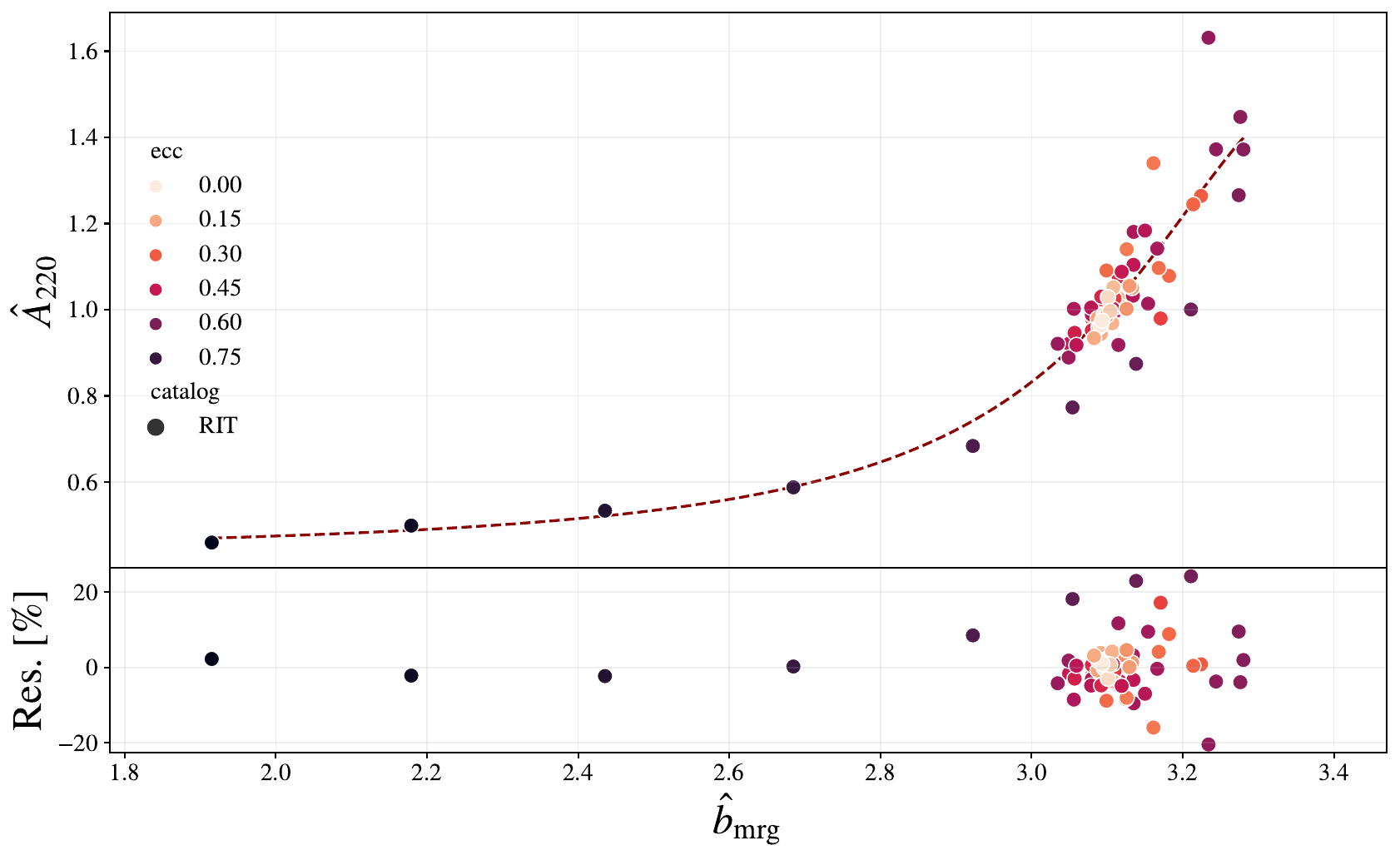}
    \includegraphics[scale=1, width=0.45\textwidth]{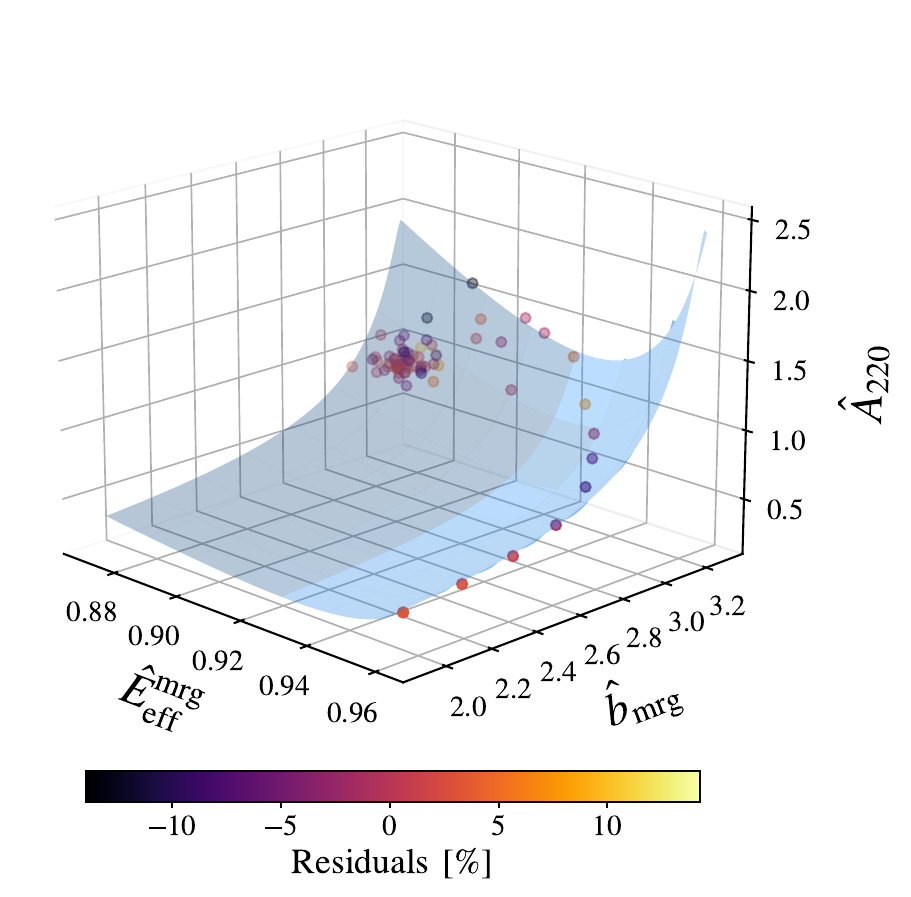}     
    \includegraphics[scale=1, width=0.45\textwidth]{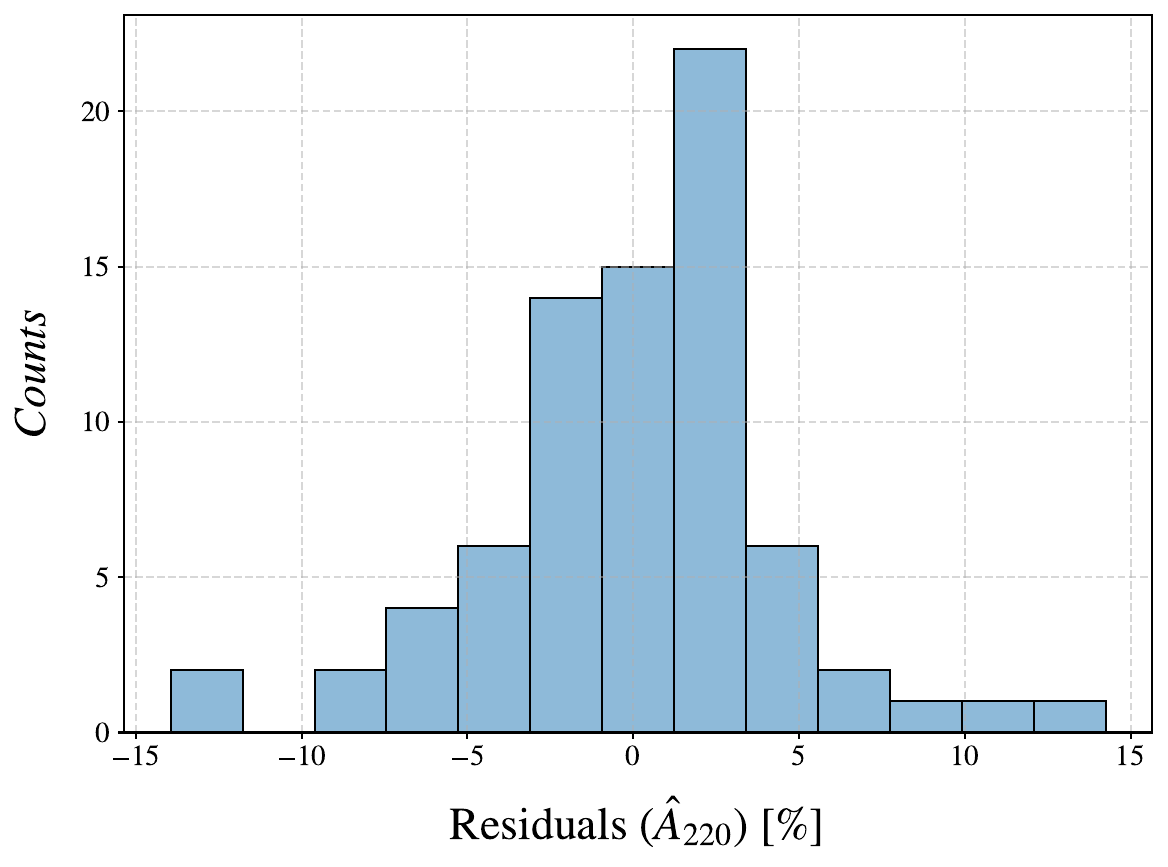}
   \caption{Equal mass case for the dominant quadrupolar mode.
   Top: fit in terms of a single variable representing the two-dimensional parameter space for noncircular initial conditions (``quasi-universal'' case). 
   The spiral-like structure in the residuals indicates the degree of quasi-universality breaking for the quantity under consideration.
   Dots are colored in terms of initial eccentricity.
   Bottom: fit in terms of the full two-degrees of freedom representing initial conditions (left); histograms of corresponding residuals (right).
   Dots are colored in terms of residuals.}
   \label{fig:2Dfit_equal_mass}
\end{figure*}

\begin{figure*}[thbp]
\centering
    \includegraphics[scale=1, width=1.0\textwidth]{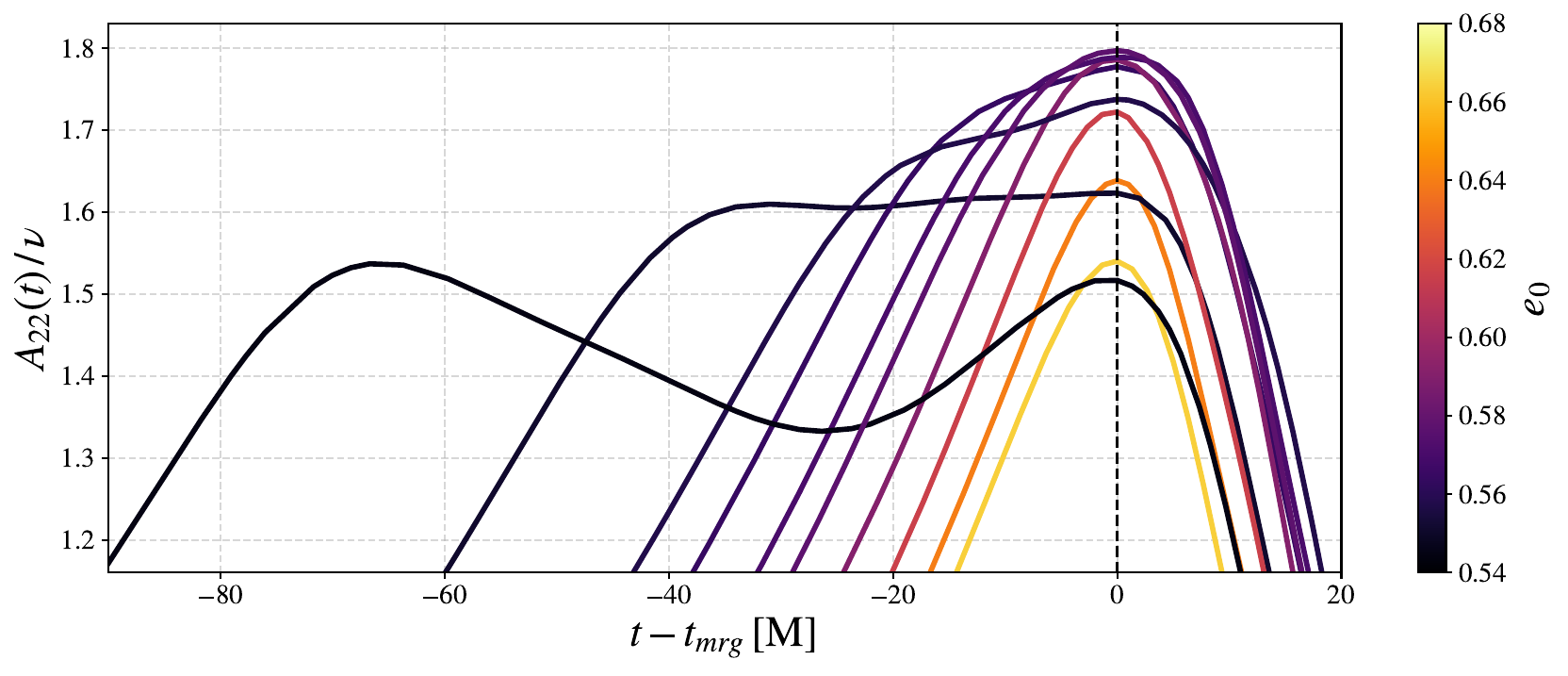}
   \caption{Zoom-in on the multipolar mode amplitude time-dependence (rescaled by the mass ratio), as a function of the eccentricity.
   The figure includes all the nonspinning equal-mass simulations with initial eccentricity in the $e_0 = [0.54, 0.68]$ range, to highlight the transition regime between an ``orbital-type'' and an ``infall-type'' phenomenology.}
   \label{fig:Amp_ecc}
\end{figure*}

\subsection{Fit construction}

To construct the complex amplitudes global fit, the output of the previous single-binary fitting procedure is gathered in a single dataset, and a model against the set of variables $(\nu, \, \bmrg, \, \Eeffmrg)$ is built.
The fitting procedure is similar to Ref.~\cite{Carullo:2023kvj}, recapped below together with the changes specific to the new quantities considered here.

First, instead of modeling the amplitudes (phases) directly, their ratio (difference) is modeled compared to the quasi-circular case.
This choice allows to scale any leading order dependence (e.g. on $\nu$ or $\Eeffmrg$) 
observed independently of the binary orbital class.
It also allows to obtain a factorized (additive) form for our fits, which can be straightforwardly applied on top of quasi-circular values~\cite{Kamaretsos:2011um,Kamaretsos:2012bs,London:2014cma,London:2018gaq,Forteza:2022tgq,Cheung:2023vki}, maintaining the accuracy of the quasi-circular limit where more numerical simulations of higher accuracy are available.
Specifically, for $(\ell, m, n) = [(2,2,0), (3,3,0), (2,1,0), (4,4,0), (3,2,0)]$, the following quantities are modeled:

\be
\hat{A}_{\ell m n} = A_{\ell m n}/A^{qc}_{\ell m n} \, ,
\ee

and 
\be 
\hat{\phi}_{\ell m n} = 2 \cdot \phi_{\ell m n} - m \cdot \phi_{220} - \phi^{qc}_{\ell m n} \, .
\ee

Quasi-circular values obtained in Ref.~\cite{Cheung:2023vki} are used, since this most recent fit relies on the highly-accurate SXS catalogue~\cite{Boyle:2019kee}.
The additional subtraction of the $m \cdot \phi_{220}$ factor serves to eliminate the dependence on the arbitrary phase of the specific simulation, similarly to Refs.~\cite{London:2018nxs, Forteza:2022tgq, Cheung:2023vki}.
This is not a restrictive assumption, since it amounts to a global phase that can always be reabsorbed in the spherical-harmonic $e^{i m \phi}$ factor, which explains the $m$ prefactor applied in the subtraction.
The above definition automatically imposes $\phi_{220}=0$, consistently with the conventions of Ref.~\cite{Cheung:2023vki}, which is followed here.
The phases $\hat{\phi}_{\ell m n}$ are also further unwrapped, i.e. the transformation $\hat{\phi}_{\ell m n} \rightarrow \hat{\phi}_{\ell m n} + 2n\pi$ is applied for the smallest possible integer $n$ that ensures continuity.
Above, a previous unwrapping stage was applied, ensuring continuity only at the level of a single binary extraction.
This second unwrapping stage is instead required to ensure continuity at the level of the entire dataset.
Since now the phase values to unwrap are multi-dimensional (as opposed to the one-dimensional unwrapping stage applied above), the standard \texttt{numpy.unwrap} function~\cite{numpy} serves to enforce continuity for most regions of the parameter space, but not for the entire set of datapoints of all modes.
This is because the \texttt{numpy} unwrapping algorithm is constructed for one-dimensional functions.
More sophisticated multi-dimensional unwrapping procedures should in principle have been applied, but this step was found to be unnecessary, as continuity could be enforced by manual addition of $2\pi$ factors for the $(2,1,0), (4,4,0), (3,2,0)$ modes in regions of the parameter space easy identified in three-dimensional plots\,\footnote{A multi-dimensional unwrapping algorithm is available e.g. at \href{https://github.com/geggo/phase-unwrap}{\color{teal} github.com/geggo/phase-unwrap}.}.
The resulting phases are then continuous, although now not in general within the $[0,2\pi]$ interval, which is still allowed given the phases periodicity.

To fit the above quantities, rational polynomials are used

\be\label{eq:template_2D}
    \tilde{Y} = \prod_{i=1}^{N} \, \tilde{Y}_0 \, \left( \frac{1 + p_{1,i} \, Q_i + p_{2,i} \, Q_i^2}{1 + p_{3,i} \, Q_i + p_{4,i} \, Q_i^2} \right) \,,
\ee

where $(Y_0, \, p_{k,i}) \in \mathbb{R}$, $N$ is the number of fitting variables considered.
In practise, $N=2$ for ``quasi-universal'' fits for which $Q_i$ runs on $(\nu, \, \bmrg)$, while $N=3$ in the cases where $Q_i$ runs on $(\nu, \, \bmrg, \, \Eeffmrg)$.
The $Y$ variable denotes either of the fit quantities: $(\hat{A}_{\ell m n}, \, \hat{\phi}_{\ell m n})$ for $(\ell, m,n) = [(2,2,0), (3,3,0), (2,1,0), (4,4,0), (3,2,0)]$.
This class of functions was found to be sufficiently simple and flexible to capture 
the structure of the complex amplitudes considered here, similarly to what observed in Ref.~\cite{Carullo:2023kvj}.

A posteriori checks of anomalous fit evaluation values were performed over a cubic dense grid constructed using the mininum and maximum values of the fitting variables $(\nu, \, \bmrg, \, \Eeffmrg)$.
Values larger than a (conservative) factor of 5 compared the maximum numerical datapoint were searched for, finding none.
This check is especially useful in the $N=3$ fitting case, where a immediate visual assessment of the fit is not possible, and serves to ensure no fit blow up either at the boundaries or at the poles of the rational functions are present.
Such discontinuities were instead found in certain cases with higher-order polynomials, which we thus avoid.
Exploration of more complex functions is left to future work, when more accurate data will be available, since: 
i) based on quasi-circular results, rational expressions are expected to be well-suited to describe these quantities; 
ii) at the moment it is not clear whether a meaningful improvement can be obtained by more complex functions, given the limited resolution of the dataset under consideration.
Keeping simple fitting functions also helps in avoiding overfitting.

Lacking reliable simulation errors, a standard non-linear least squares is used for the fit residuals minimisation.
The \texttt{scipy.optimize.least\_squares} function is adopted, bounding all the coefficients within $[-100,100]$.
Algorithm convergence was enforced by repeating the fit using 10 distinct seed values, verifying that different seeds deliver compatible residuals, and selecting the maximum likelihood value.
Seed values are drawn from a gaussian distribution $\mathcal{N}(1, 1)$.
Spot-checks of dataset outliers did not reveal any obvious issue with the specific underlying simulation giving rise to anomalous results.

Residuals for the amplitudes are defined by $\Delta Y \equiv (Y - Y_{\mathrm{NR}})/Y_{\mathrm{NR}}$, while for the phases it is more meaninful to report residuals as $\Delta Y \equiv (Y - Y_{\mathrm{NR}})$, to avoid discontinuities due to zero-crossings.
The obtained coefficients and a \texttt{python} implementation of the fits are publicly available at: \href{https://github.com/GCArullo/noncircular_BBH_fits}{github.com/GCArullo/noncircular\_BBH\_fits}.

\subsection{Results}

\begin{figure*}[thbp]
\centering
    \includegraphics[scale=1, width=0.43\textwidth]{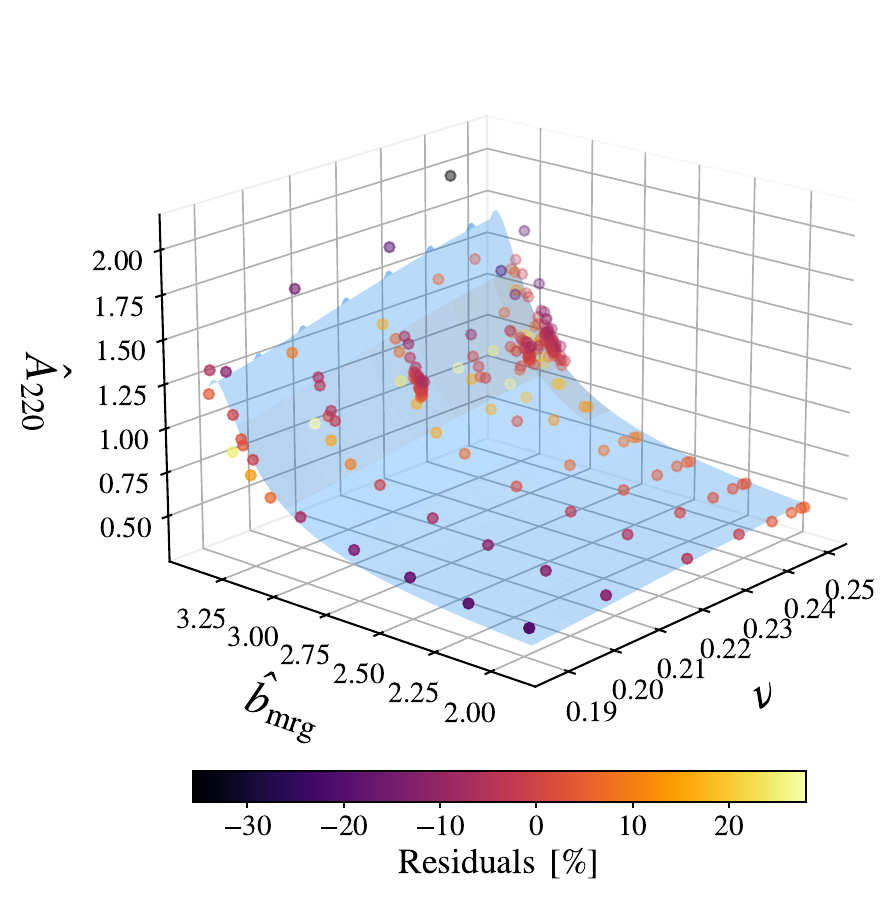}     
    \includegraphics[scale=1, width=0.43\textwidth]{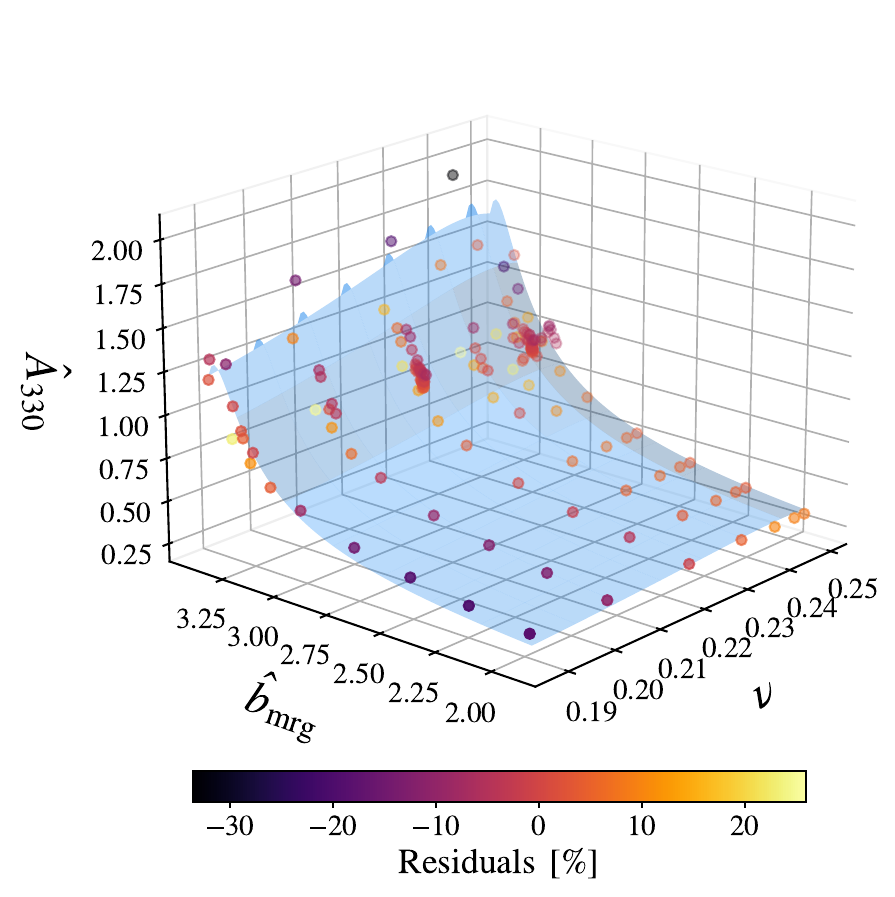}
    \includegraphics[scale=1, width=0.43\textwidth]{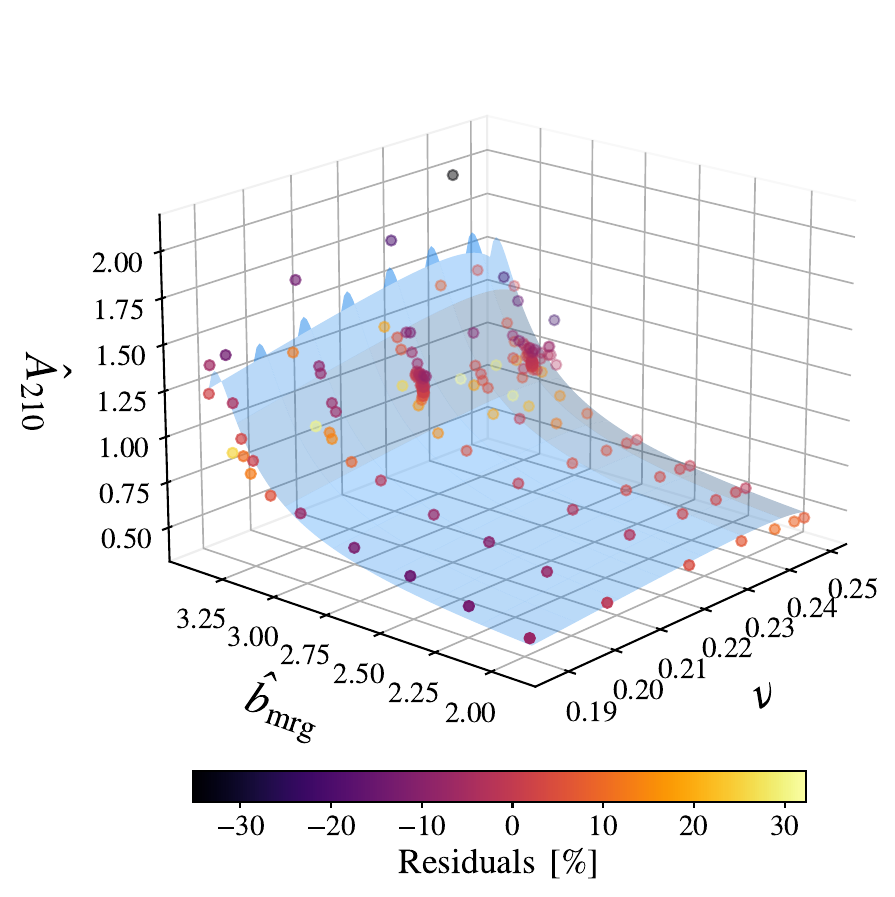}
    \includegraphics[scale=1, width=0.43\textwidth]{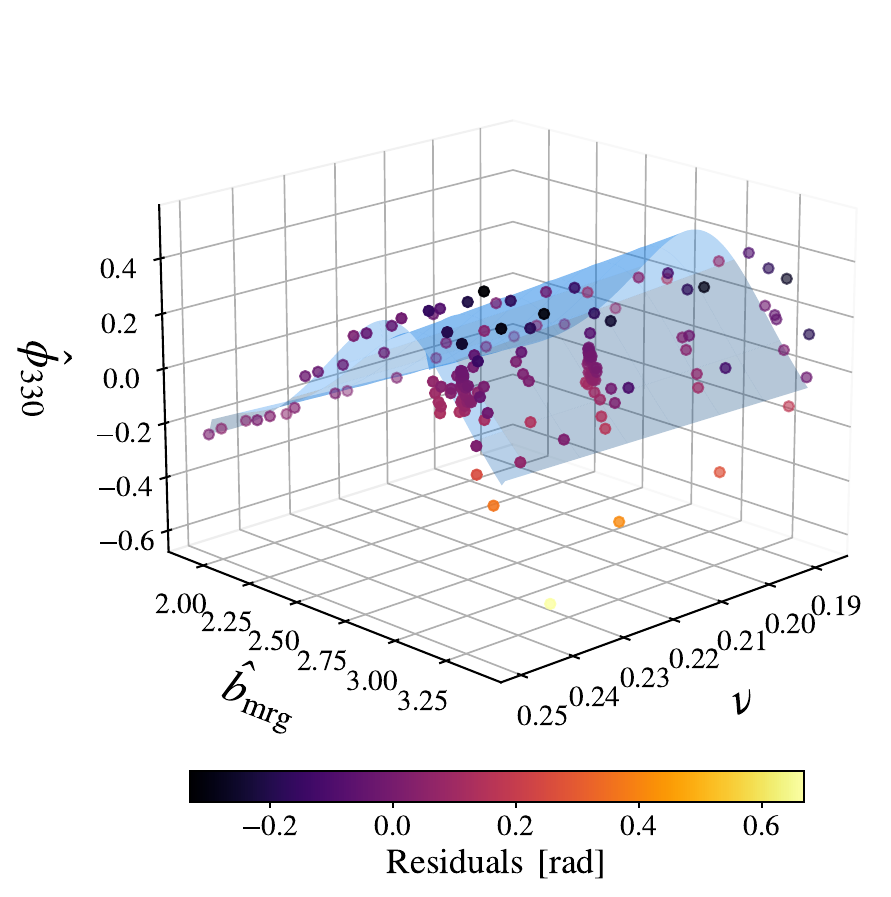}
    \includegraphics[scale=1, width=0.43\textwidth]{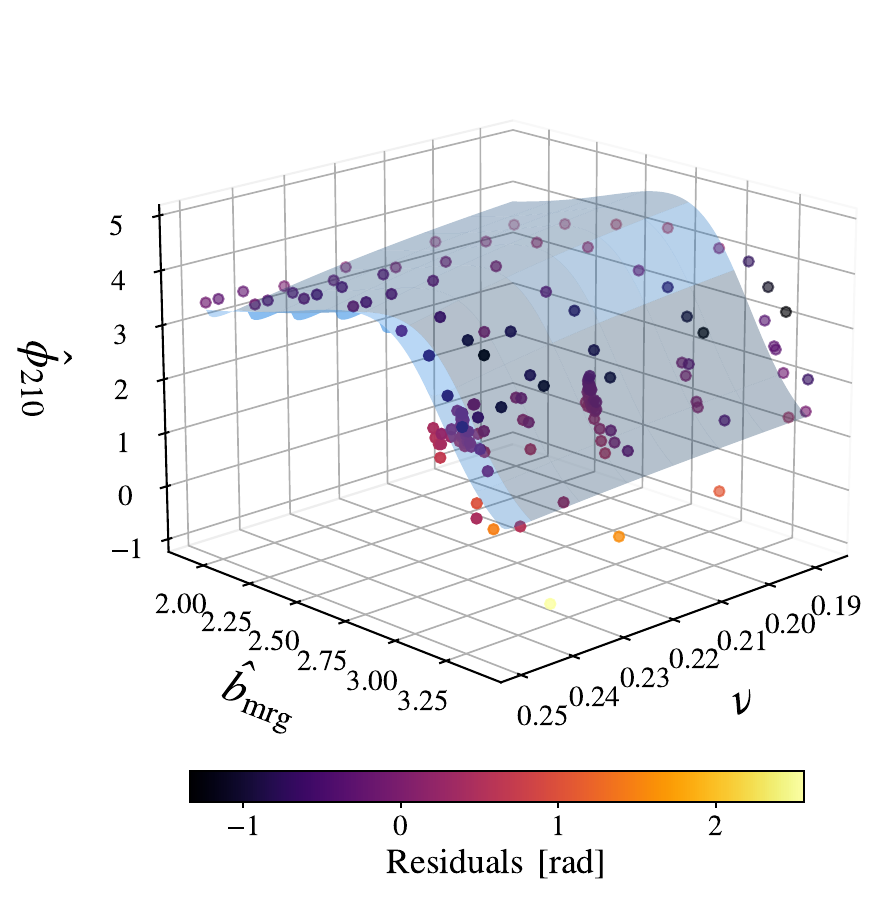}
   \caption{Unequal mass case for the dominant ringdown modes, in terms of a single variable representing initial conditions (``quasi-universal'' case) and mass ratio. 
   Datapoints scattering indicates the degree of quasi-universality breaking for the quantity under consideration.
   The wiggles and layer-crossing features, indicating a non-monotonic behaviour, are only an artifact of the 2D projection.
}
   \label{fig:2Dfit}
\end{figure*}

\begin{figure*}[thbp]
\centering
    \includegraphics[scale=1, width=0.48\textwidth]{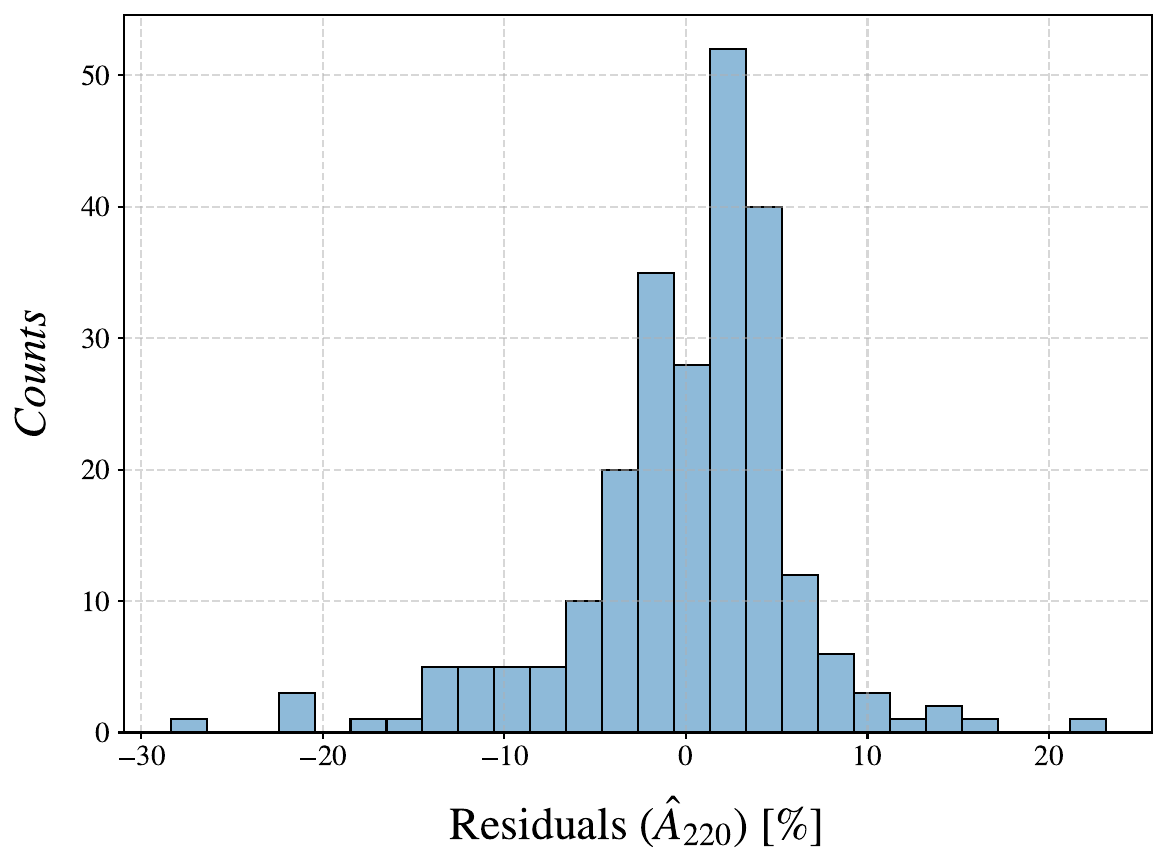} 
    \includegraphics[scale=1, width=0.48\textwidth]{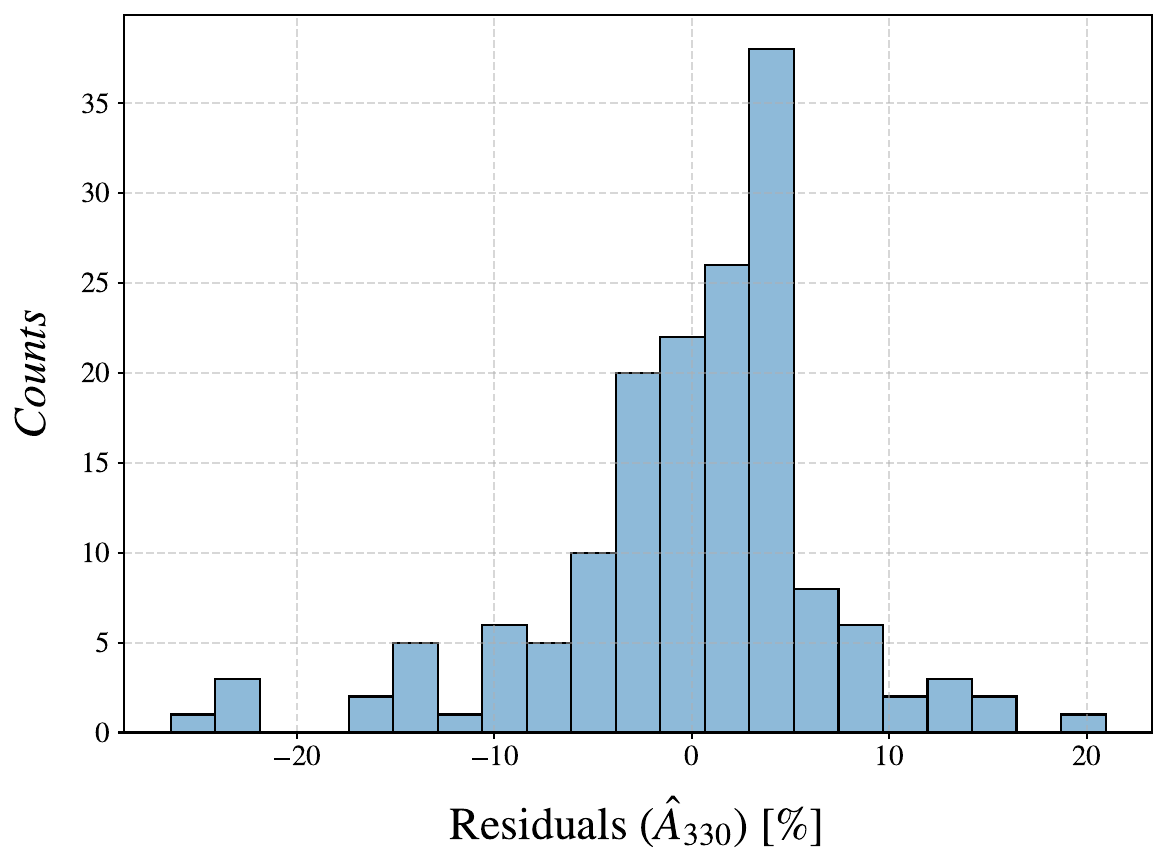}    
    \includegraphics[scale=1, width=0.48\textwidth]{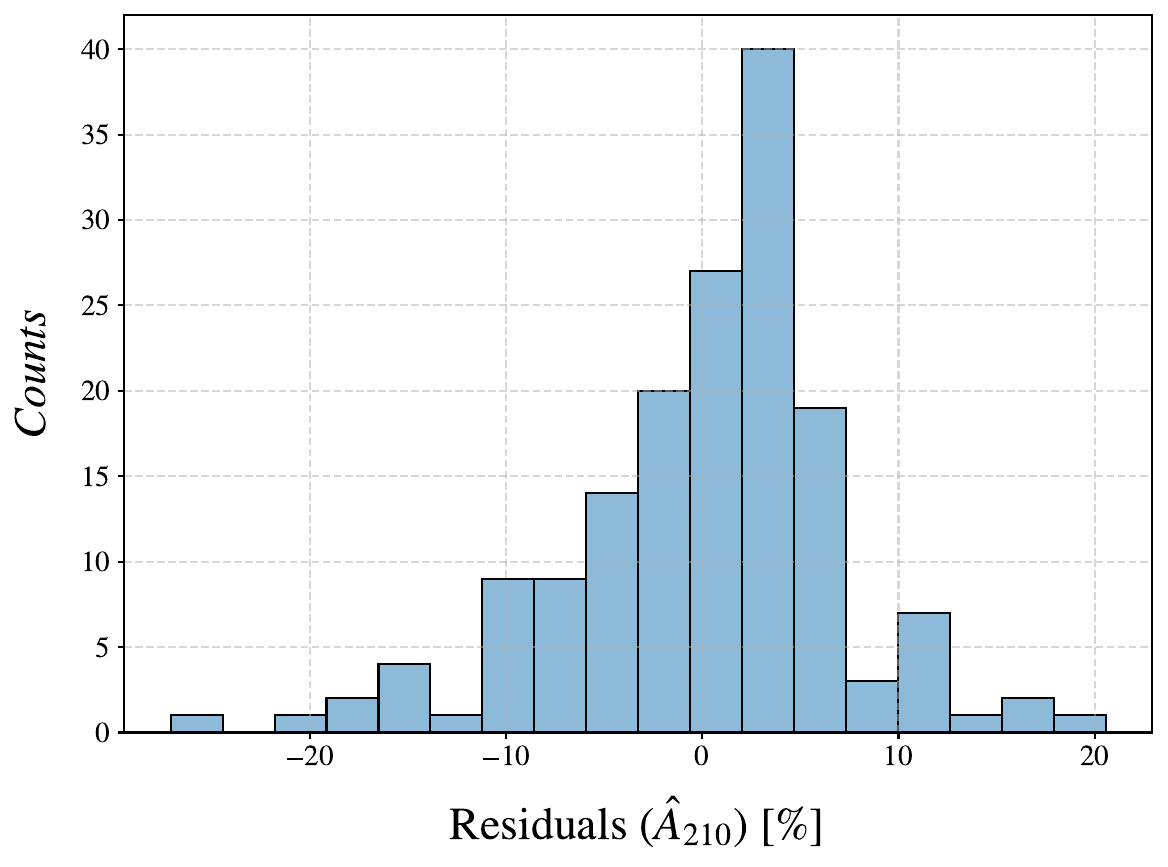}  
    \includegraphics[scale=1, width=0.48\textwidth]{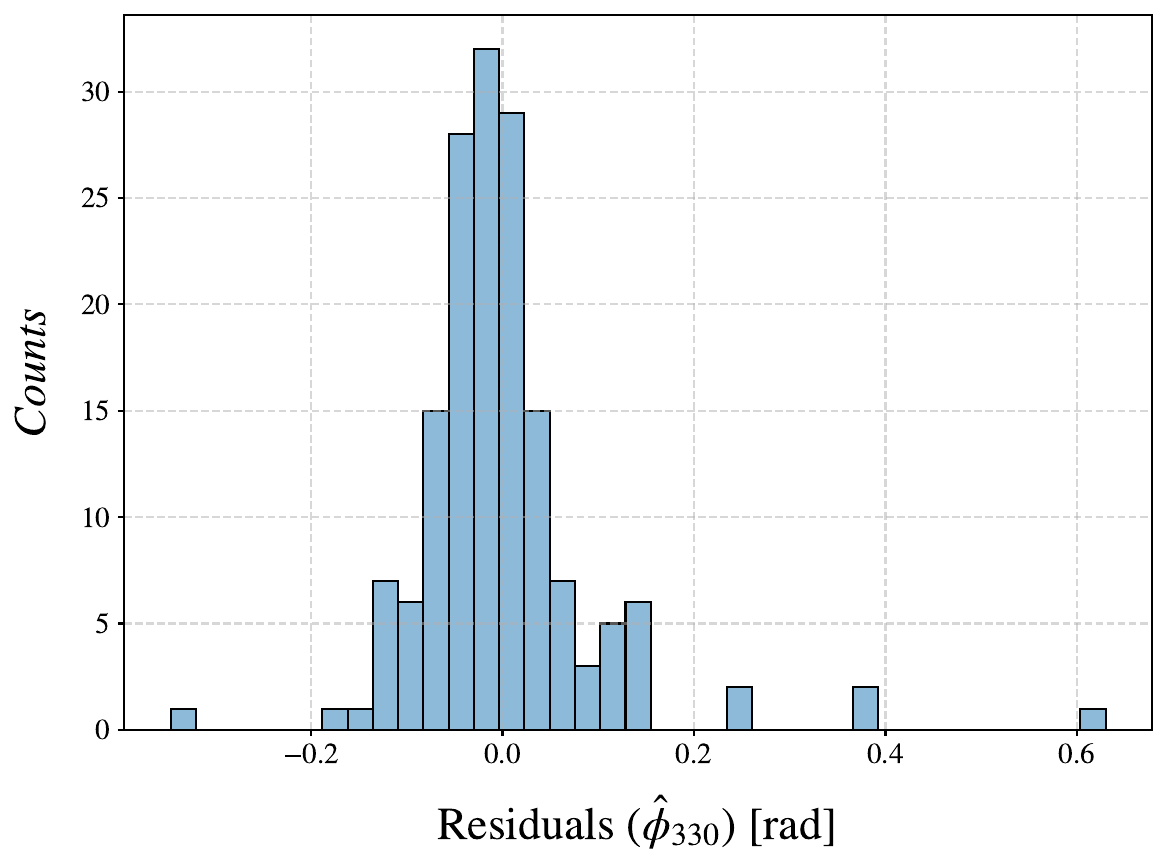}
    \includegraphics[scale=1, width=0.48\textwidth]{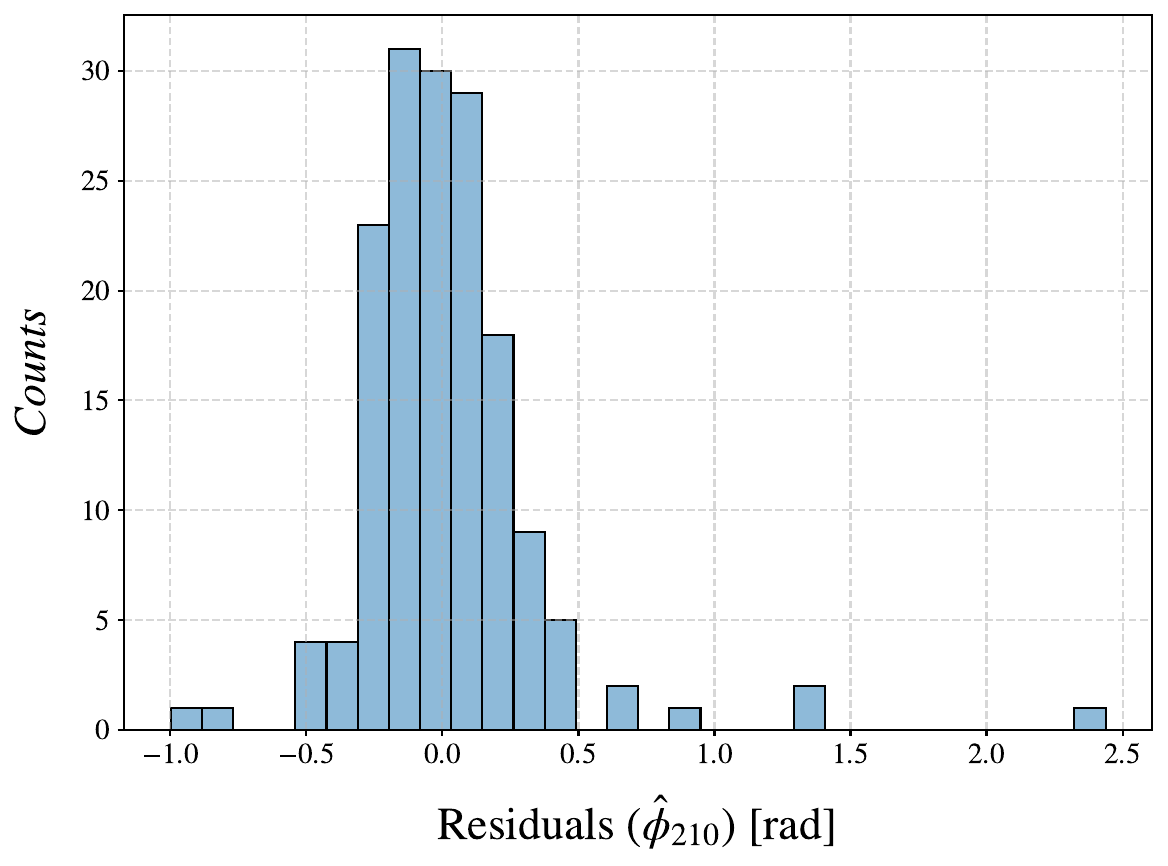}
   \caption{Residuals of the three dimensional fits, in terms of the full two-degrees of freedom representing initial conditions, and the symmetric mass ratio.}
   \label{fig:residuals_3Dfit}
\end{figure*}

The dataset under consideration, constituting the final outcome of the procedure presented in Sec.~\ref{sec:amplitude_extraction}, is displayed in Fig.~\ref{fig:3D_dataset} for the most excited modes and in Fig.~\ref{fig:440_fits_unequal_mass} for other sub-dominant modes.
Here, the value of the amplitude and phases is shown in terms of the two non-circular parameters $(\bmrg, \Eeffmrg)$ and the mass ratio $\nu$.
A non-monotonic but overall smooth structure can be observed, with the exception of certain datapoints displaying what seem to be anomalous values.
Indeed, as will be discussed in more detail below, a few simulations appear as outliers with respect to the overall dataset, which might signal undetected simulation inaccuracies not captured by our data-quality filter based on balance laws.

\begin{figure*}[thbp]
\centering
    \includegraphics[scale=1, width=0.48\textwidth]{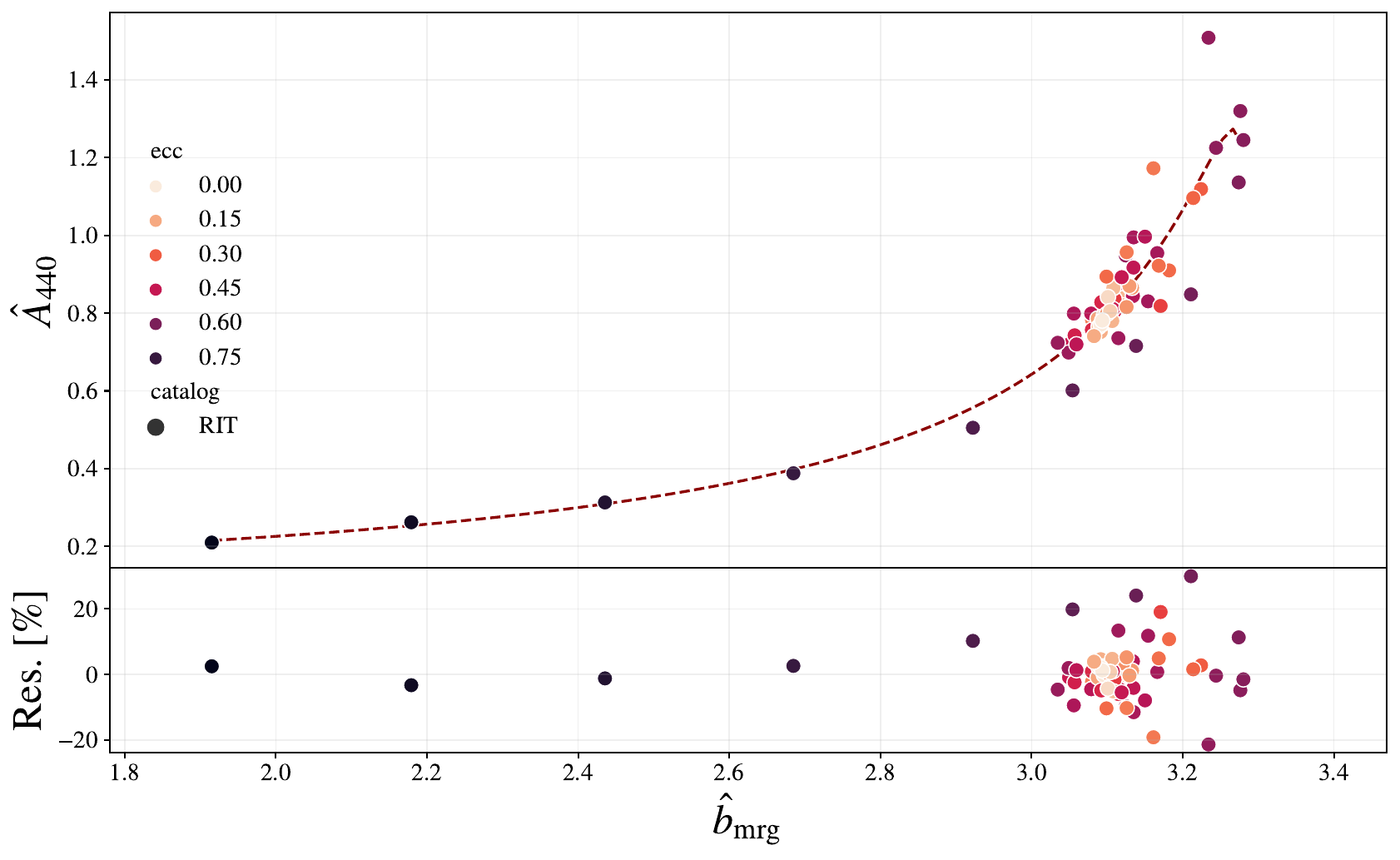}
    \includegraphics[scale=1, width=0.48\textwidth]{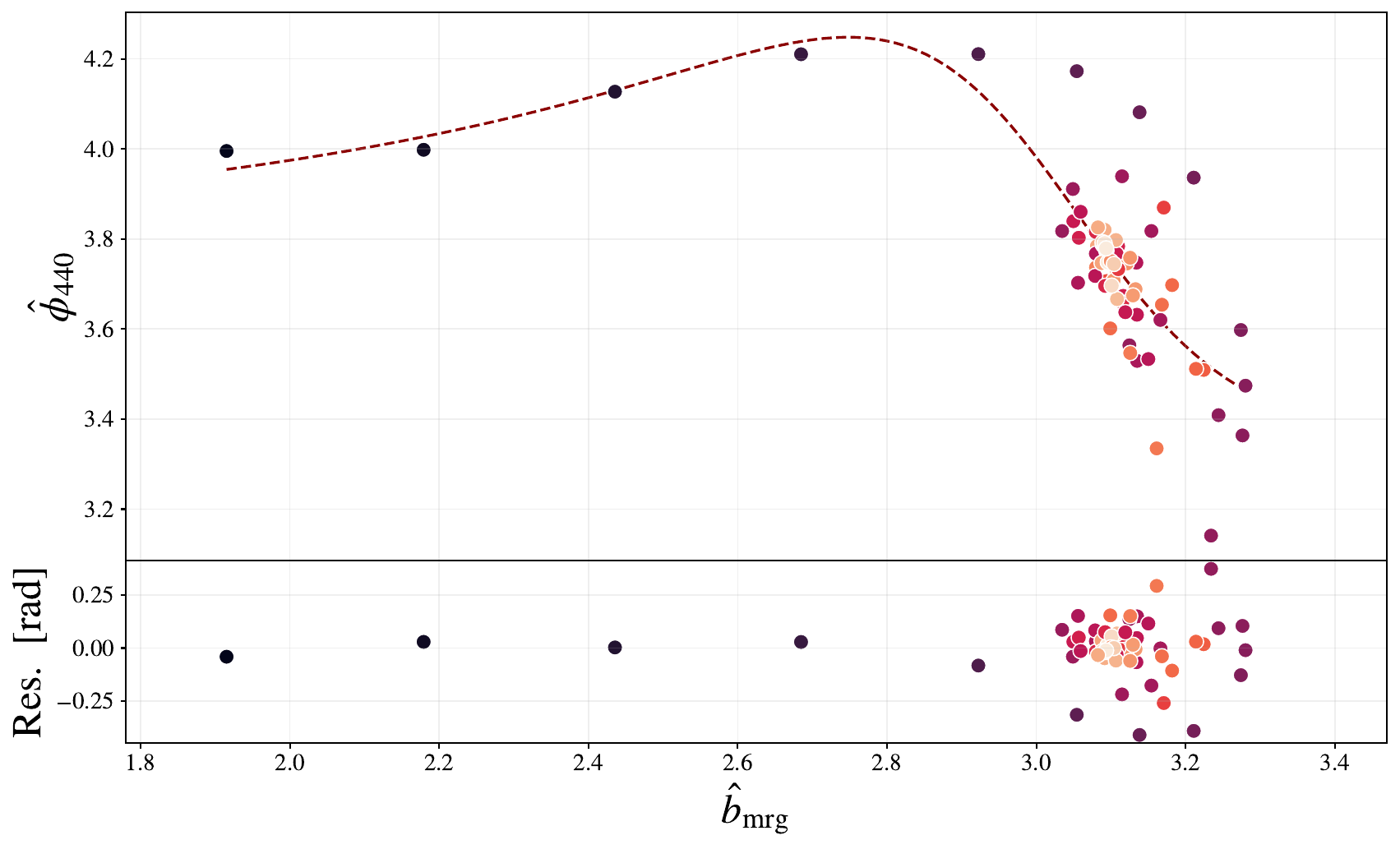}
    \includegraphics[scale=1, width=0.45\textwidth]{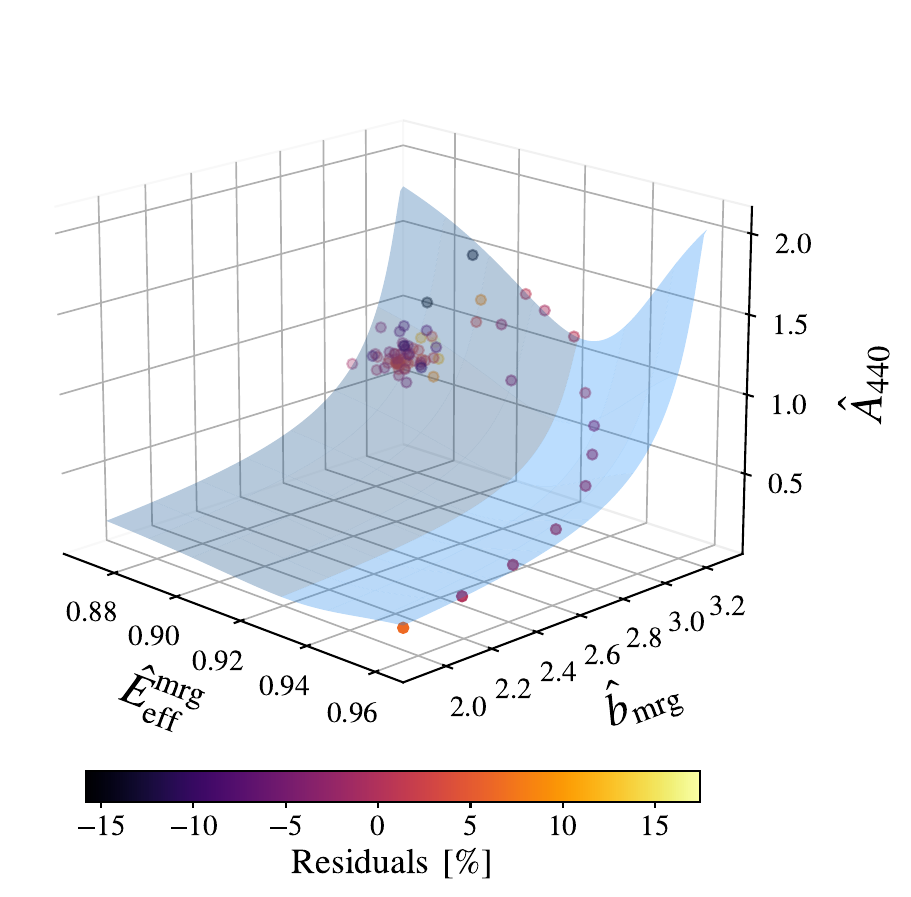}
    \includegraphics[scale=1, width=0.45\textwidth]{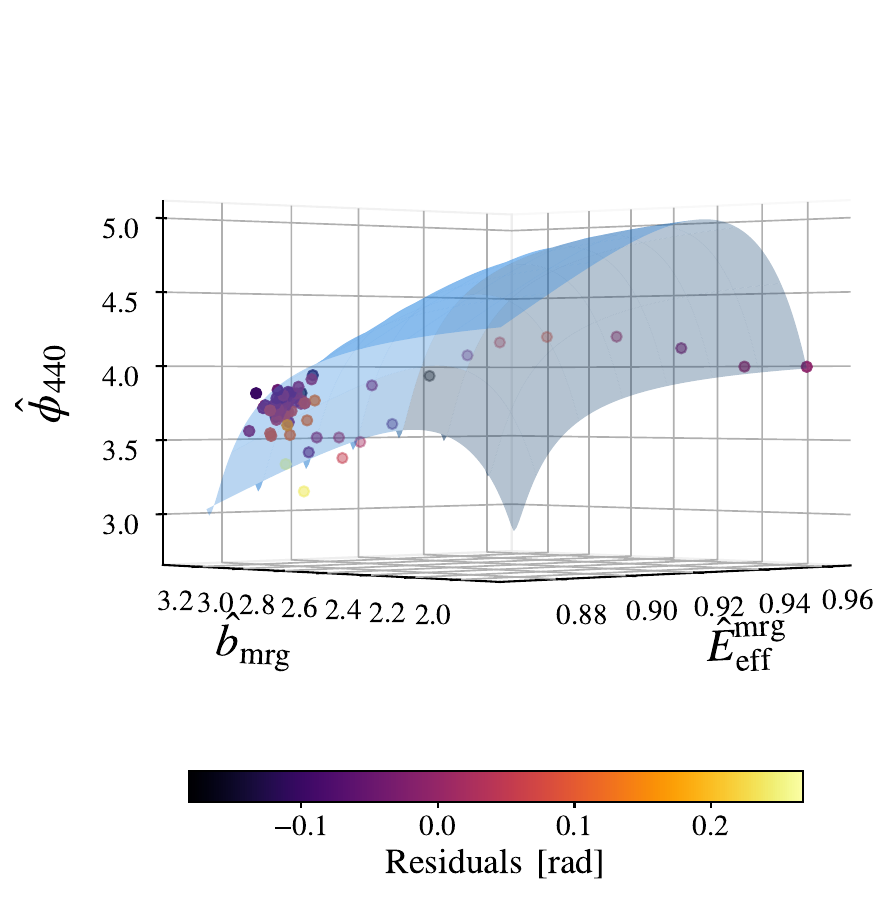}   
    \includegraphics[scale=1, width=0.45\textwidth]{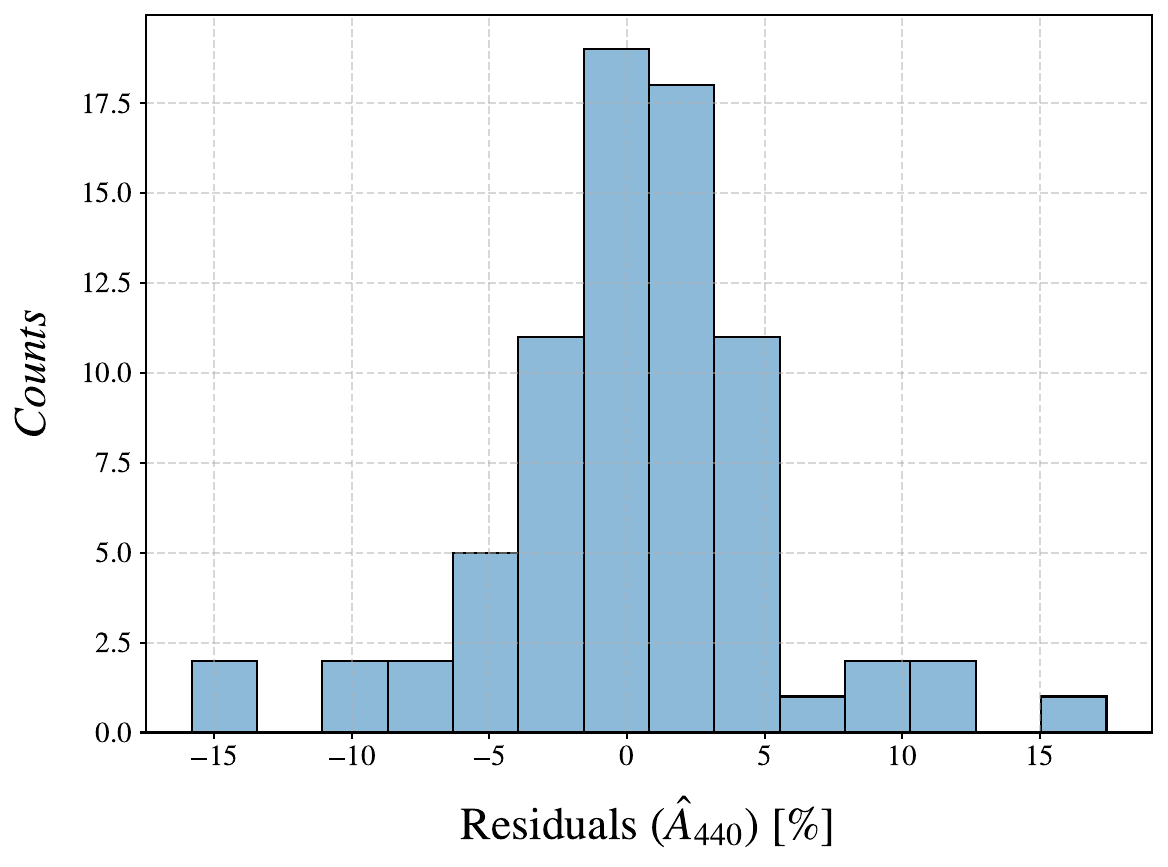}
    \includegraphics[scale=1, width=0.45\textwidth]{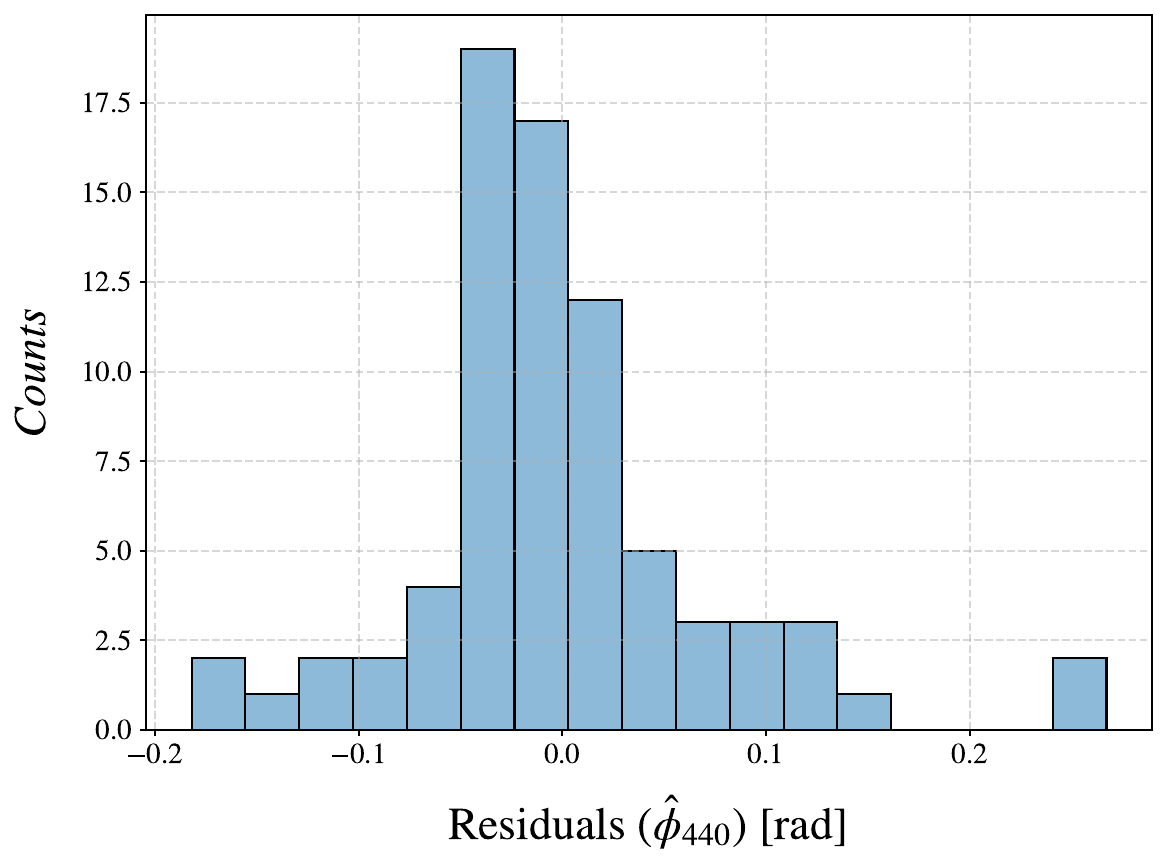}
   \caption{Equal mass case for the dominant hexadecapolar mode.
   Top: fit in terms of a single variable representing noncircular initial conditions (``quasi-universal'' case). 
   Dots are colored by values of initial eccentricity.
   The spiral-like structure in the residuals indicates the degree of quasi-universality breaking for the quantity under consideration.
   Middle: fit in terms of the full two-degrees of freedom representing initial conditions.
   Dots are colored in terms of residuals.
   Bottom: histograms of residuals corresponding to the two-dimensional case.}
   \label{fig:2Dfit_equal_mass_440}
\end{figure*}

\begin{figure*}[thbp]
\centering
    \includegraphics[scale=1, width=0.48\textwidth]{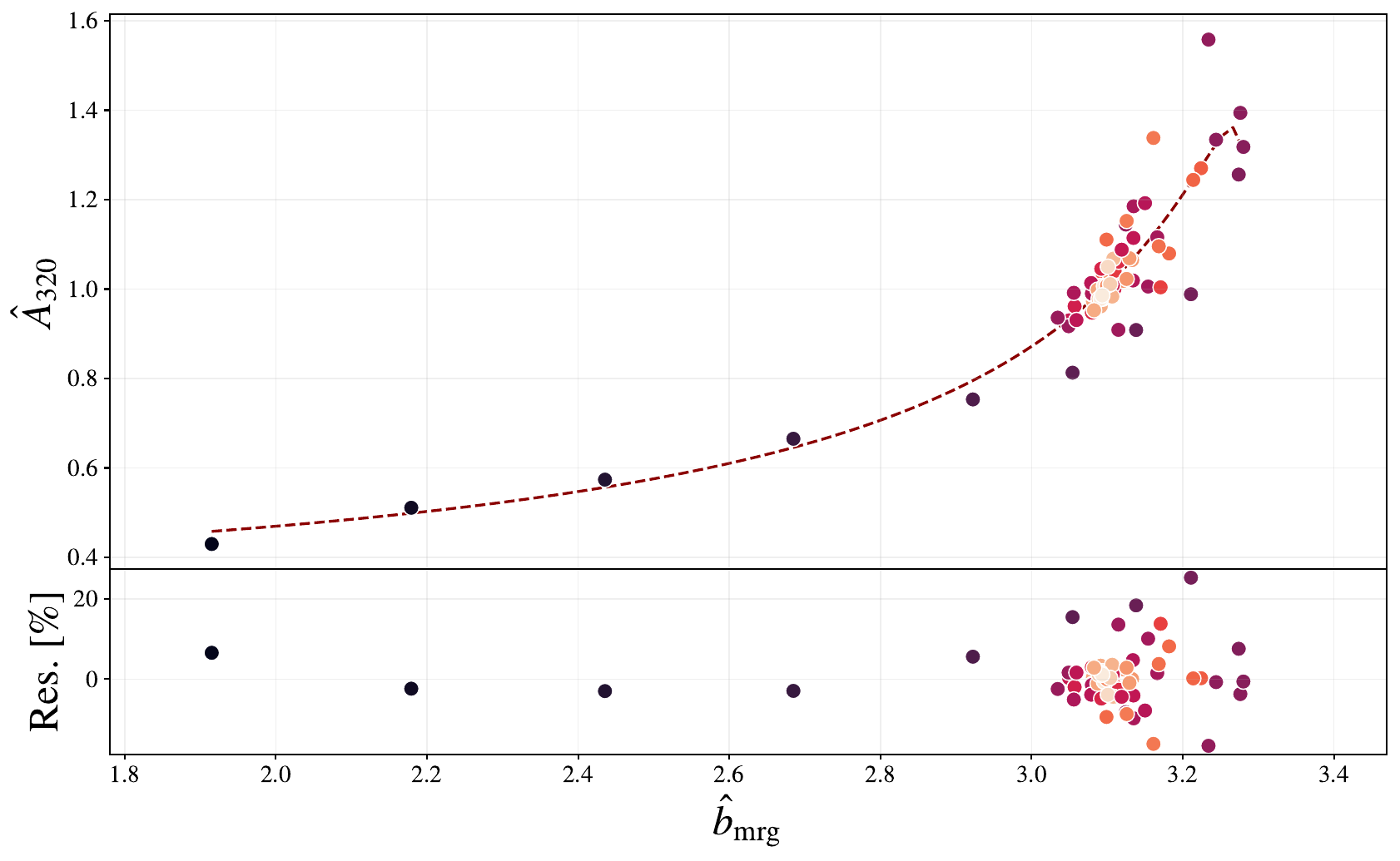}
    \includegraphics[scale=1, width=0.48\textwidth]{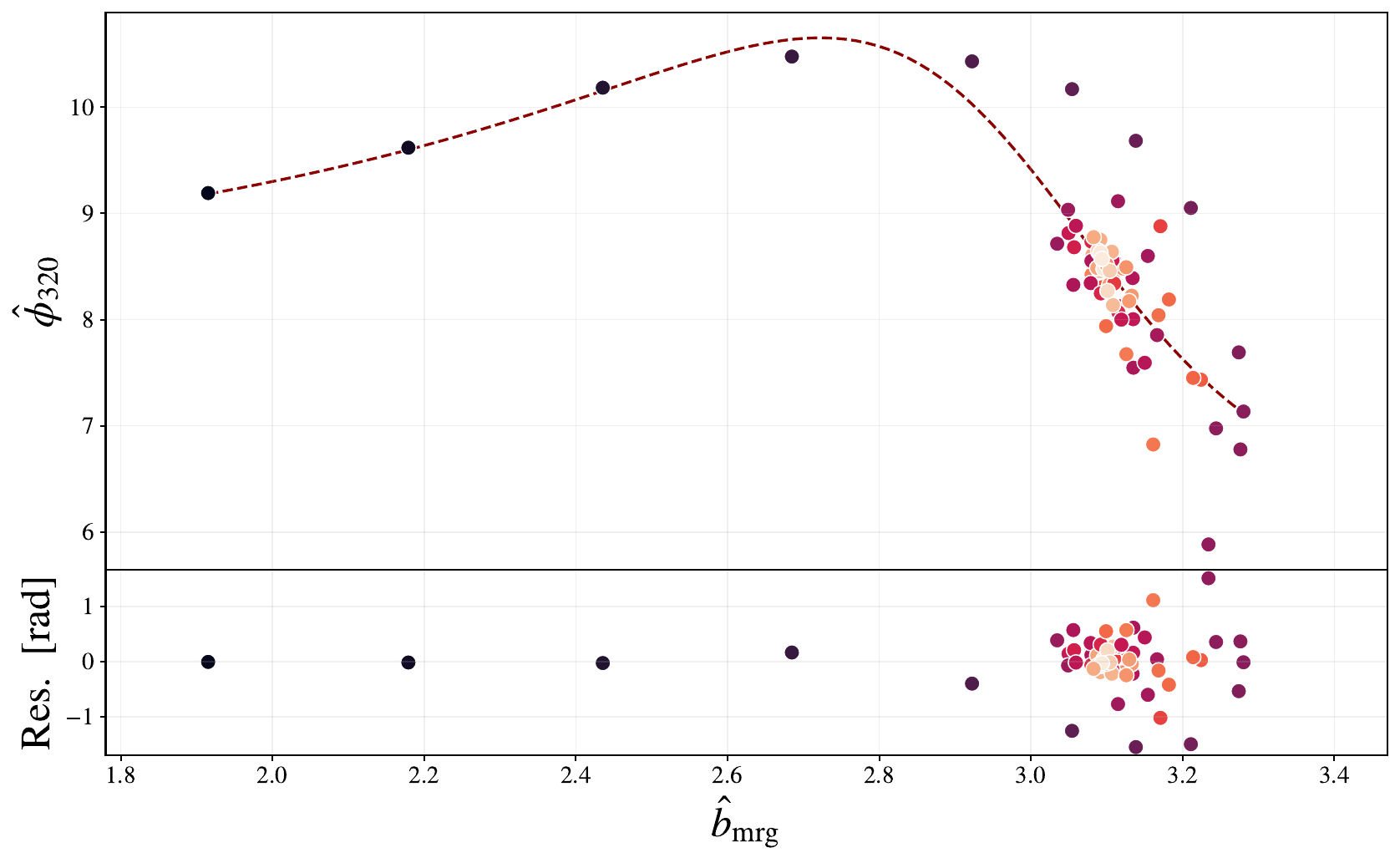}
    \includegraphics[scale=1, width=0.45\textwidth]{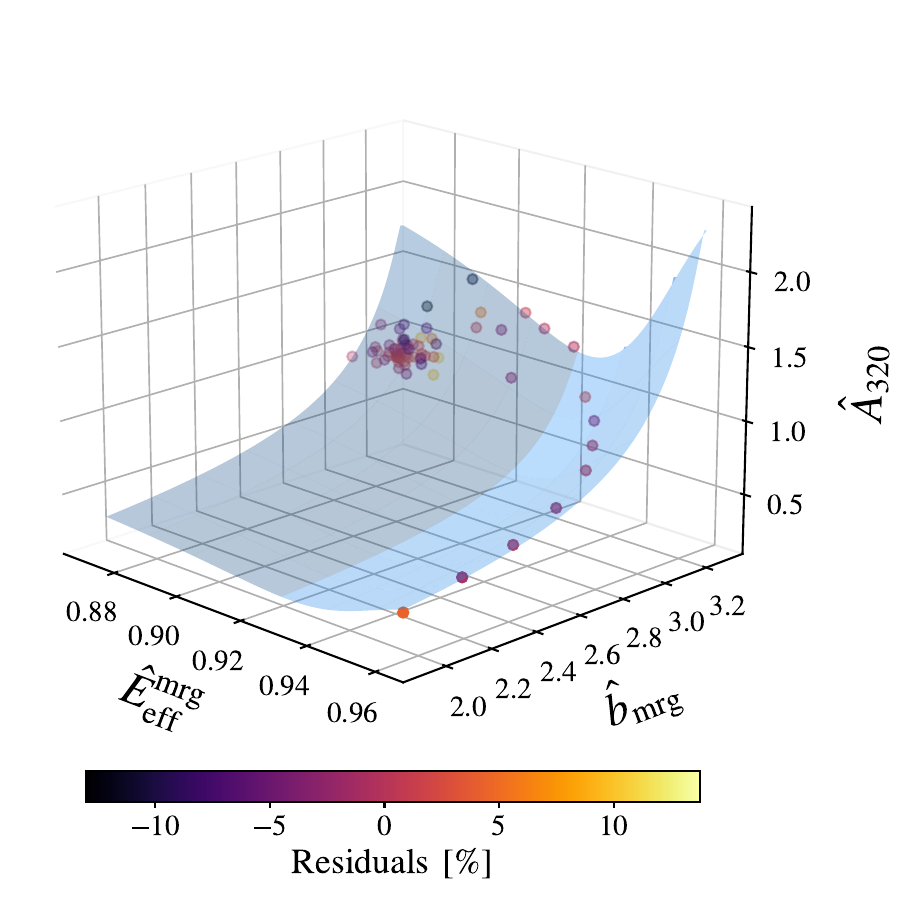}
    \includegraphics[scale=1, width=0.45\textwidth]{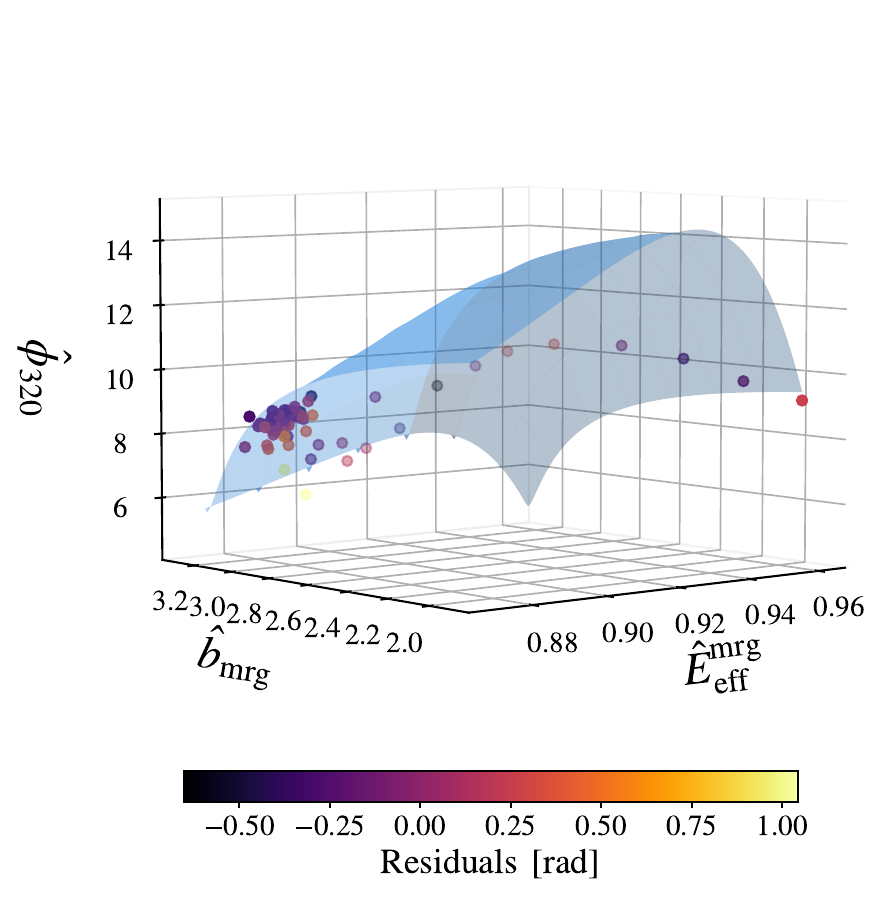}   
    \includegraphics[scale=1, width=0.45\textwidth]{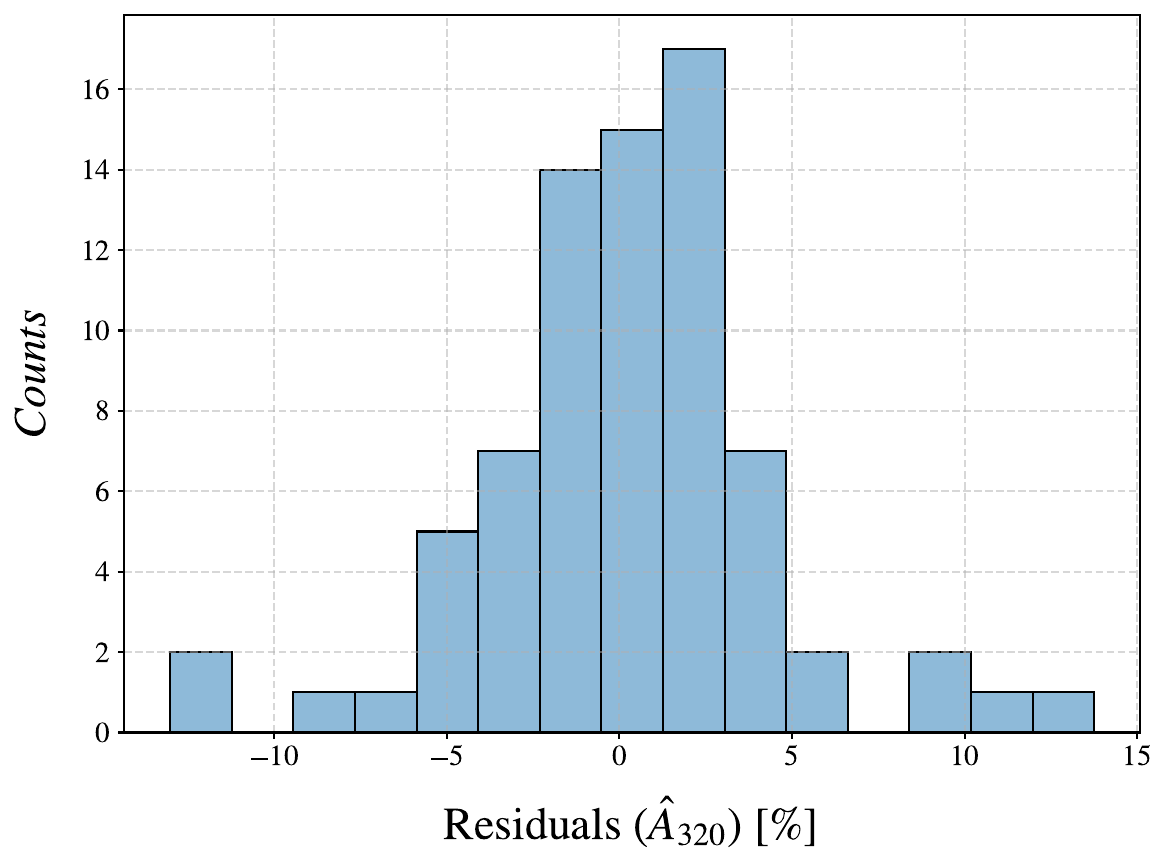}
    \includegraphics[scale=1, width=0.45\textwidth]{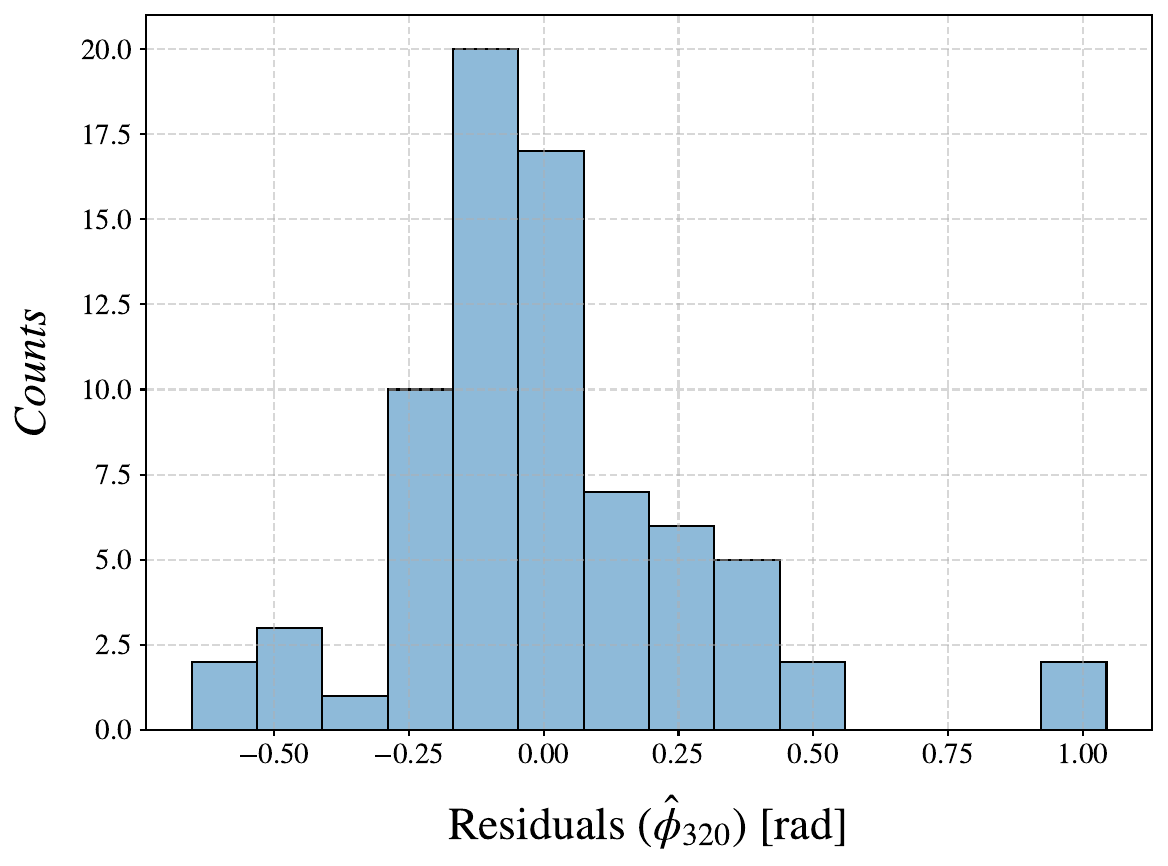}
   \caption{Equal mass case for the dominant mode receiving mode-mixing, $(\ell,m)=(3,2)$.
   Top: fit in terms of a single variable representing noncircular initial conditions (``quasi-universal'' case). 
   Dots are colored by values of initial eccentricity.
   The spiral-like structure in the residuals indicates the degree of quasi-universality breaking for the quantity under consideration.
   Middle: fit in terms of the full two-degrees of freedom representing initial conditions.
   Dots are colored in terms of residuals.
   Bottom: histograms of residuals corresponding to the two-dimensional case.}
   \label{fig:2Dfit_equal_mass_320}
\end{figure*}

\begin{figure*}[thbp]
\centering
    \includegraphics[scale=1, width=0.44\textwidth]{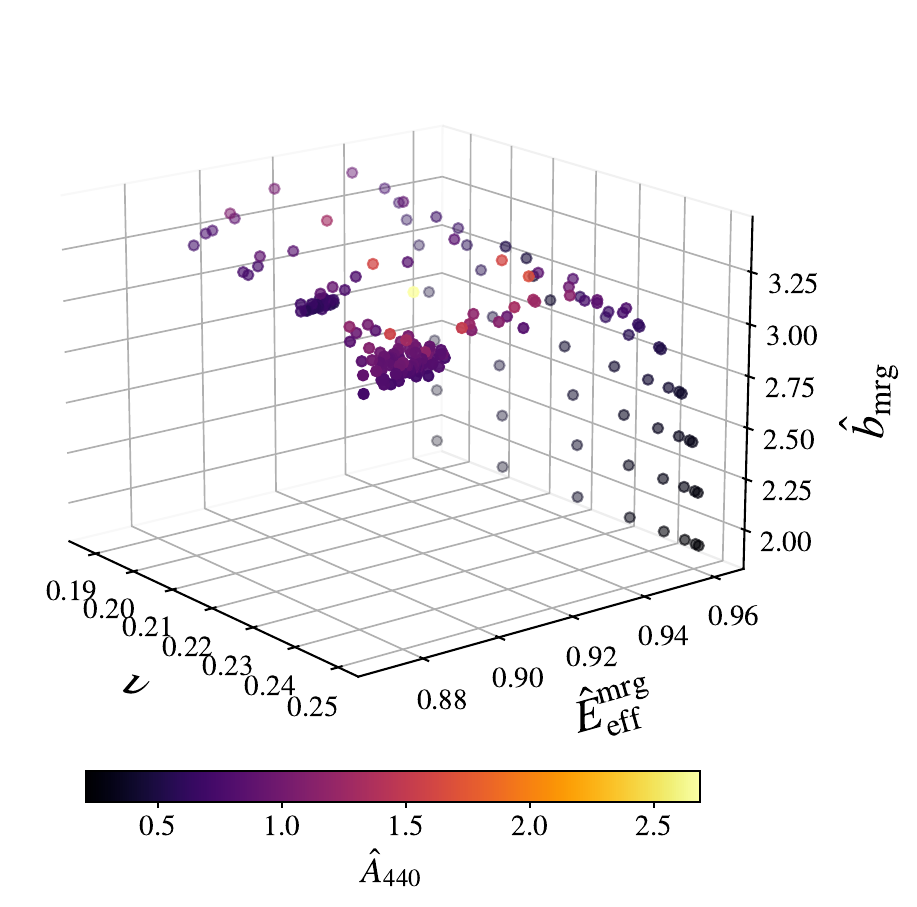}
    \includegraphics[scale=1, width=0.44\textwidth]{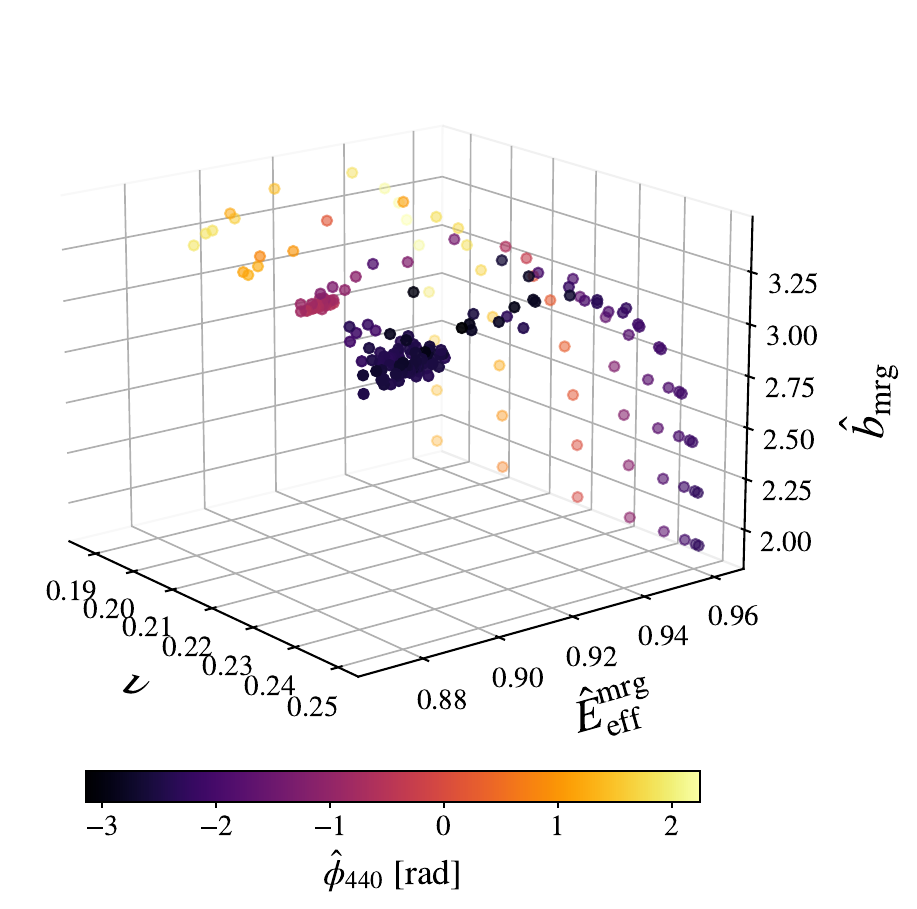}
    \includegraphics[scale=1, width=0.44\textwidth]{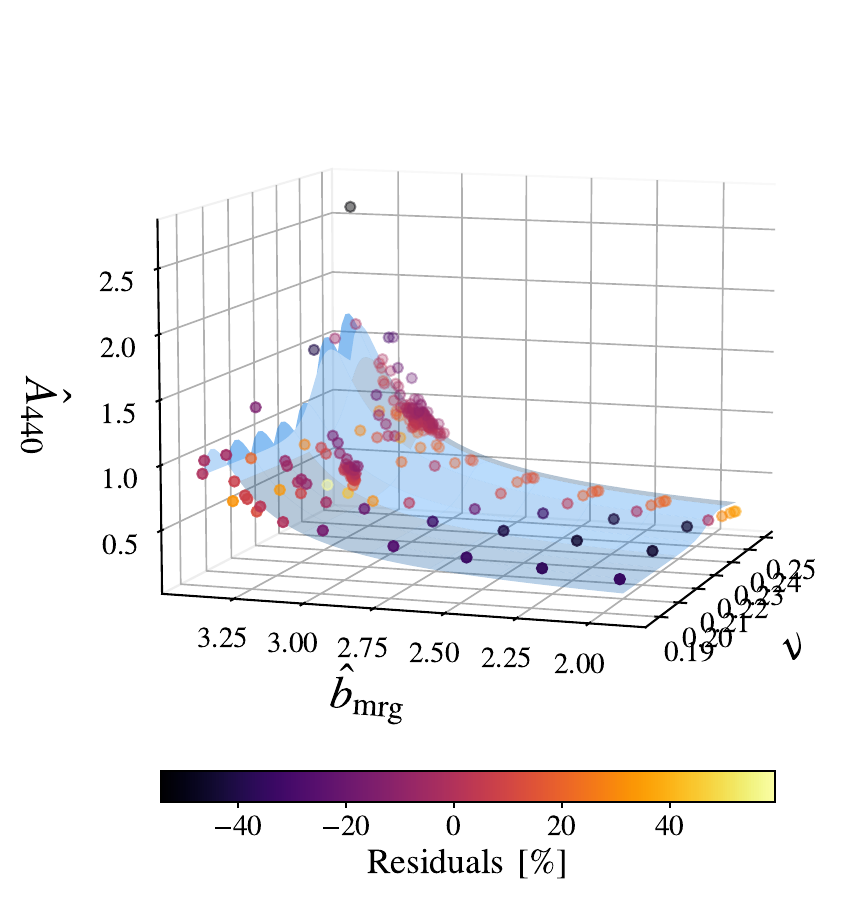}
    \includegraphics[scale=1, width=0.44\textwidth]{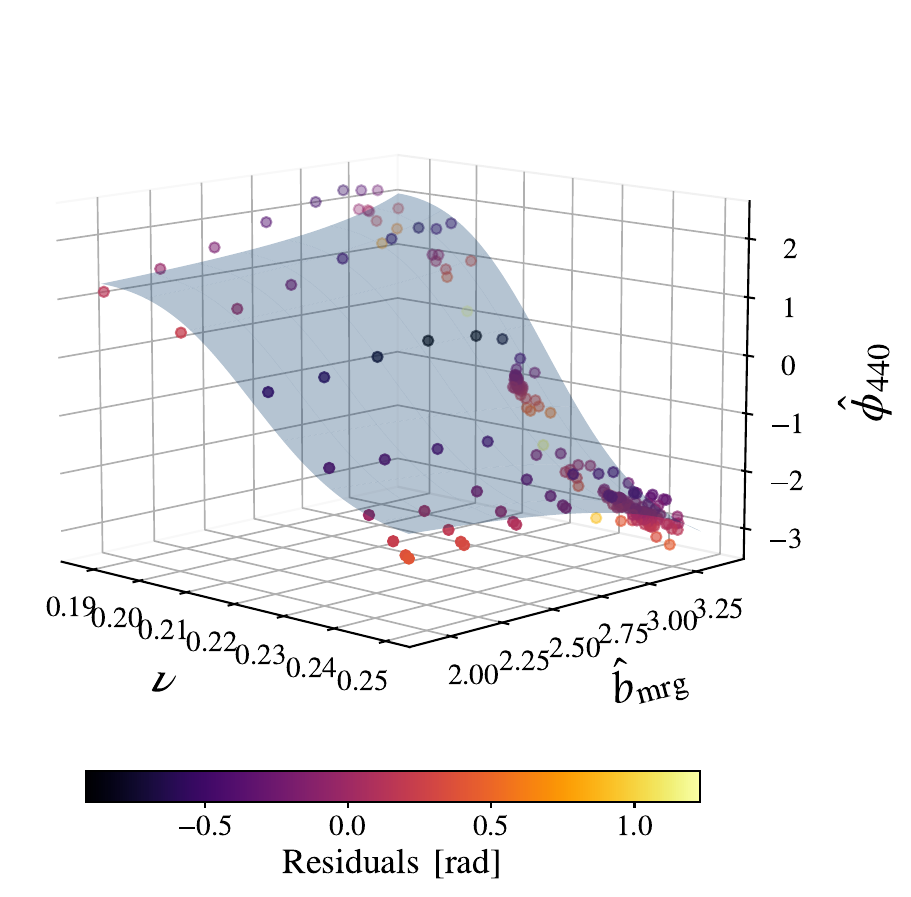}
    \includegraphics[scale=1, width=0.44\textwidth]{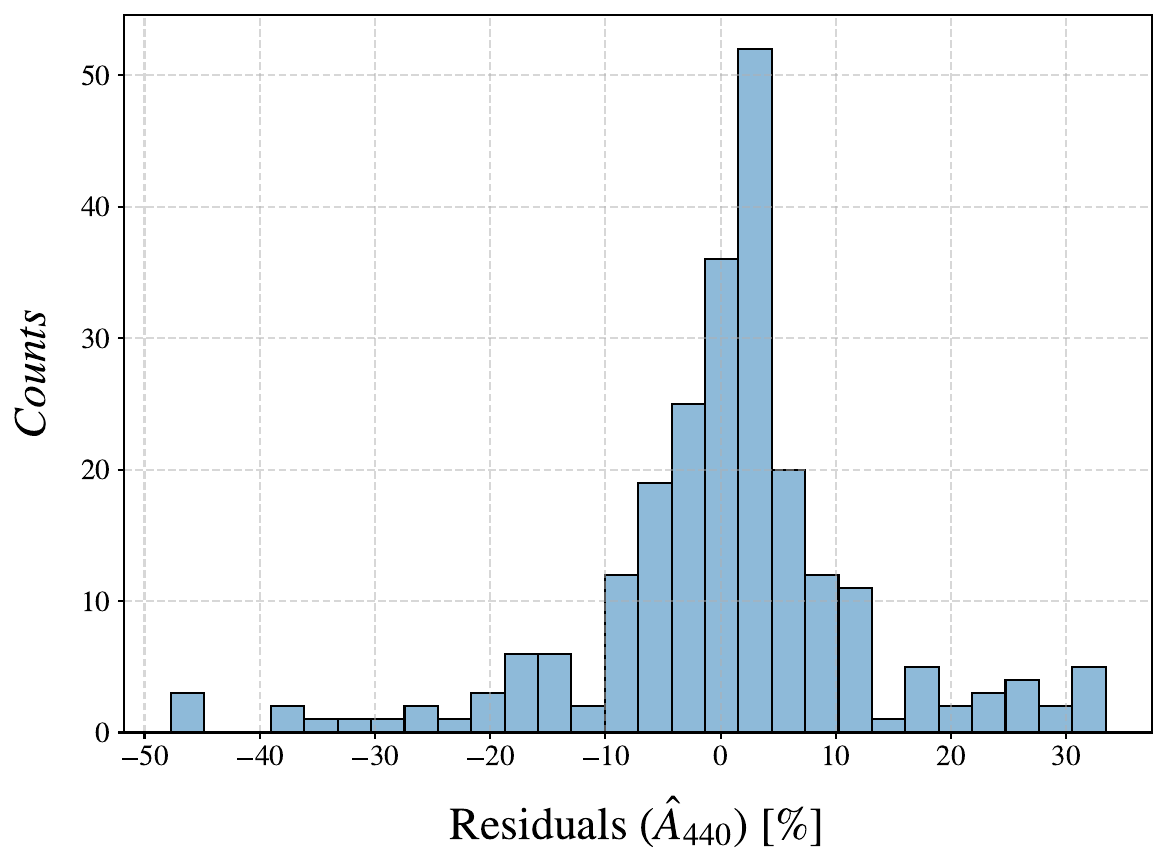}
    \includegraphics[scale=1, width=0.44\textwidth]{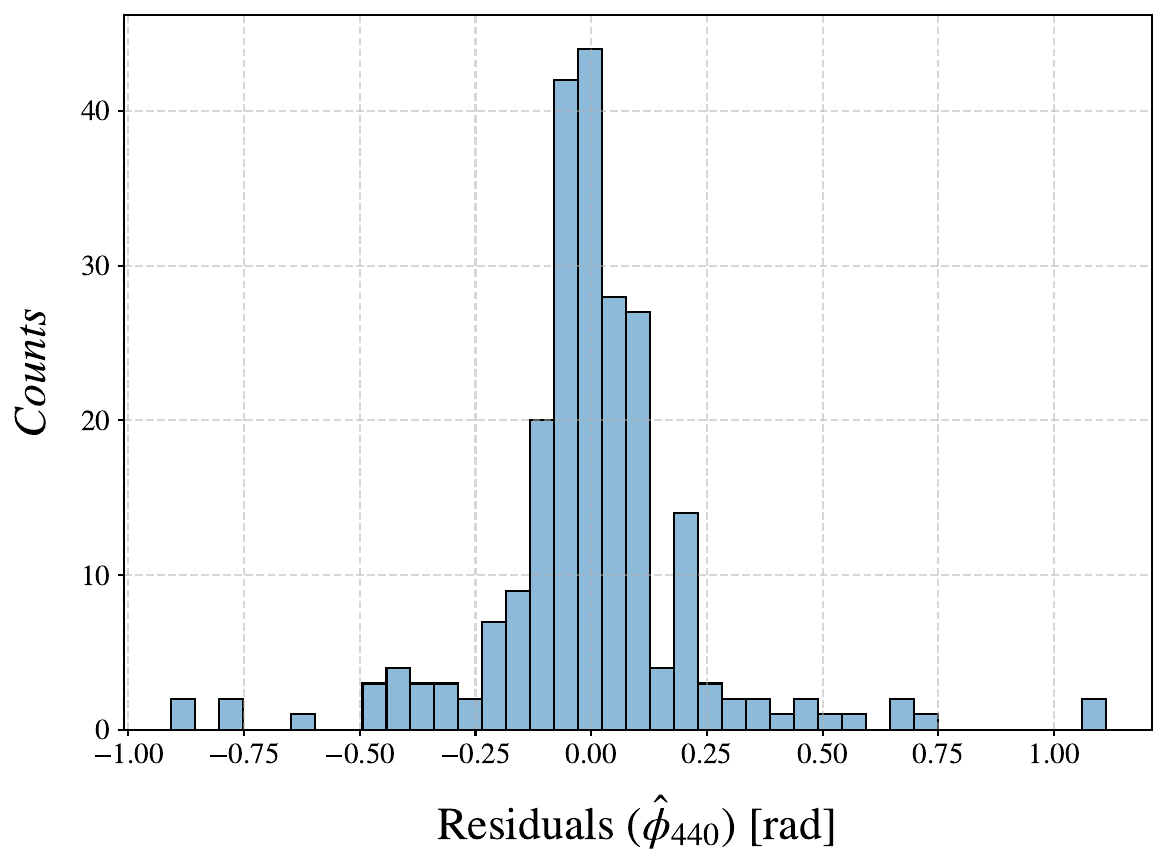}
   \caption{Parameter space (top), two-dimensional fits (middle) and three-dimensional fits (bottom) for the $(4,4,0)$ amplitude and phase parameters.
   The apparent spiky behaviour in the left central plot is a visualisation artifacts due to the surface curvature.}
   \label{fig:440_fits_unequal_mass}
\end{figure*}

Results for the equal-mass case are displayed in Fig.~\ref{fig:2Dfit_equal_mass} for the dominant quadrupolar mode.
The top panel reports on the one-dimensional fit in terms of a single noncircular variable $\bmrg$, which was found to be highly effective in modeling merger-remnant quantities in Ref.~\cite{Carullo:2023kvj}.
There, quasi-universality held at the level of a few percent for most of the cases considered in Ref.~\cite{Carullo:2023kvj}.
Here, a spiral-like structure in the residuals indicates the degree of such universality violation, and points to the need of adding a second variable to the fit.
The two-dimensional case, now in terms of both $\Eeffmrg$ and $\bmrg$ is displayed in the bottom panel.
The two-dimensional fit diplays a smooth hyperboloidal structure, and residuals decrease below $10\%$, with the exception of a few outliers.
Specifically, $92\%$ of the simulations in the dataset have residuals smaller than $5\%$, and $97\%$ below $10\%$.
Similar considerations apply to the dominant hexadecapolar mode and to the dominant mode receiving mode-mixing, $(\ell,m)=(3,2)$, as displayed in Figs.~\ref{fig:2Dfit_equal_mass_440},\ref{fig:2Dfit_equal_mass_320}.
Odd-$m$ modes are zero by symmetry in the equal-mass case~\cite{Blanchet:2013haa}.
The amplitude fit accuracy remains similar to the one of the quadrupolar mode.
For these modes we can also fit for the phase parameter (in the (2,2,0) case, the phase is zero by construction), and similar fits give residuals below $\rm 0.2 \, rad$ (except for a single case) for the $(4,4,0)$ mode, while the $(3,2,0)$ case displays a few outliers above $\rm 0.5 \, rad$.
Visual inspection of the underlying  waveforms suggest the (3,2,0) outliers are driven by systematic errors in the numerical data, and point to the need of improved numerical simulations.
Except for the few cases just discussed, the equal mass fits for the noncircular correction factors are generally comparable in accuracy to existing quasi-circular models, typically around a few percent in amplitude and a few deciradians in phase~\cite{Forteza:2022tgq}.

An important feature of Fig.~\ref{fig:2Dfit_equal_mass} deserves further discussion.
The top panel shows how the amplitude is not a monotonic function of the initial eccentricity (colorbar), unlike the impact parameter at merger (x-axis).
This is due to the non-trivial dynamics of the last periastron before merger, at the transition between an ``orbital-type'' and an ``infall-type'' dynamics.
Such dynamics is readily captured by the impact parameter evaluated around the plunge-merger stage, but not by the initial eccentricity.
This behaviour is analogous to what was observed in Ref.~\cite{Carullo:2023kvj} for the merger amplitude, and such transitional behaviour had been predicted by the model of Ref.~\cite{Nagar:2020xsk} and observed in full nonlinear simulations in Refs.~\cite{Gamba:2024cvy, Kankani:2024may, Albanesi:2024xus}.
Specifically, for medium values of eccentricity, a large burst of GWs is emitted on the last periastron just before merger, implying a large loss of energy from the system, hence a reduced merger (and similarly ringdown) amplitude.
When eccentricity increases towards $e_0 \simeq 0.6$, the last periastron radius moves closer and closer to the last stable orbit radius.
Eventually, the system cannot sustain orbits beyond this point, and the last periastron smoothly blends with the merger.
This smooth blending, visualised in Fig.~\ref{fig:Amp_ecc}, explains why these systems have the largest impact parameter at merger: these are the systems for which a periastron passage (with a larger impact parameter compared to e.g. the one of a quasi-circular binary at merger) \textit{becomes} the merger.
Specifically, the figure reports the waveform amplitude of the dominant multipole as a function of time, for a subset of nonspinning equal-mass simulations around the transition regime, spanned by the initial eccentricity.
When this transition happens, the merger amplitude initially increases since: 
i) less energy has been lost in previous encounters; 
ii) the merger happens during an inversion point of the orbit (i.e. the point that maximises the emission of GWs), which compounds with the increase in emission due to the two objects starting to touch each other.
A similar phenomenology can be observed in Fig.~1 of Ref.~\cite{Nagar:2020xsk}, which also gives an account of the binary dynamics behind this transition.
When the eccentricity increases even further, the system possesses less orbital angular momentum, hence the two objects plunge before an inversion (the would-be last periastron) can start taking place, reducing the merger amplitude.
This marks the final stage of the transition towards an ``infall-type'' merger behaviour, which continues smoothly (with a corresponding decrease of the merger amplitude visible in the above figure) until the eccentricity reaches unity, and the dynamics becomes a pure head-on case.
The ringdown amplitude has an identical behaviour, although the transition seems to happen for slightly lower eccentricity values.
All these complex features are instead immediately captured by the impact parameter, as already apparent from Fig.~1 of Ref.~\cite{Carullo:2023kvj}.
See Sec.~\ref{subsec:par} for a conjecture on why this is the case.

The unequal-mass case is instead reported in Fig.~\ref{fig:2Dfit} for the most excited modes, when assuming quasi-universality.
The fit still displays an overall smooth structure.
However, a few common outliers can be noted in the upper range of amplitude values, and lower range of phases values.
Additionally, all the amplitudes consistently display a small turnover for highest values of $\bmrg$.
Additional simulations at higher $\bmrg$ values will be required to assess if this constitutes a physical feature, or is instead induced by simulation inaccuracies.
As expected (see e.g. Ref.~\cite{London:2014cma}, phases display both a more complex structure and larger noisy features, similarly to what observed for the frequency parameter in Ref.~\cite{Carullo:2023kvj}.

\begin{figure*}[thbp]
\centering
    \includegraphics[scale=1, width=0.44\textwidth]{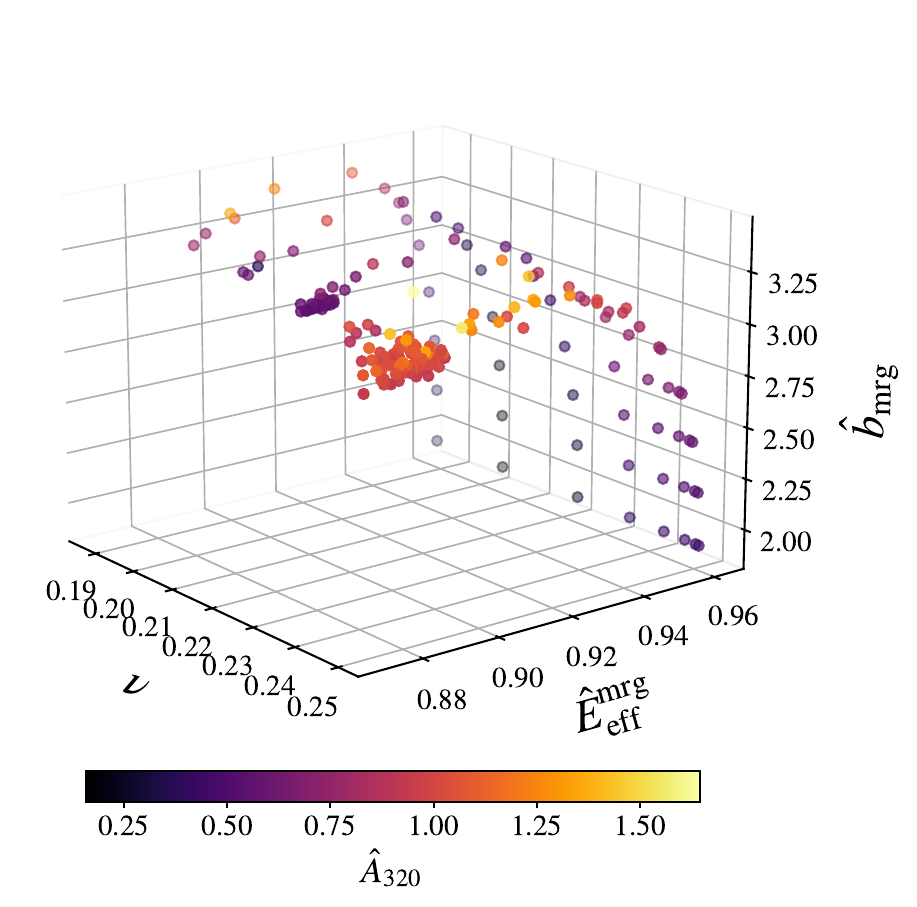}
    \includegraphics[scale=1, width=0.44\textwidth]{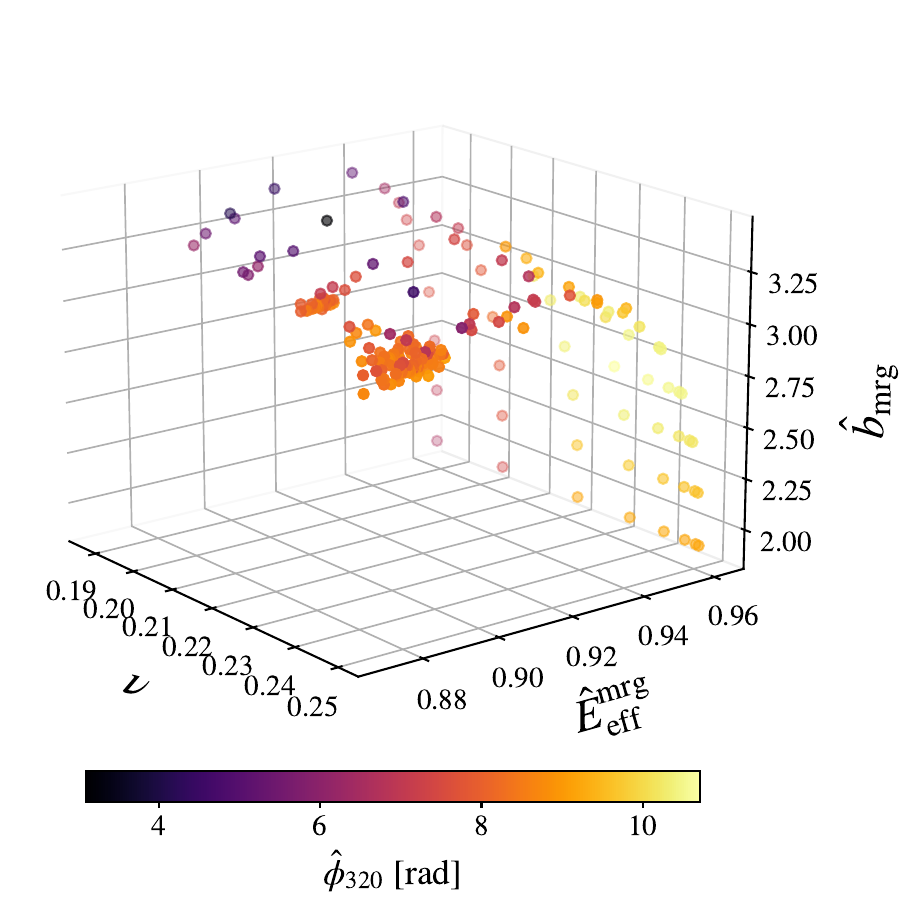}
    \includegraphics[scale=1, width=0.44\textwidth]{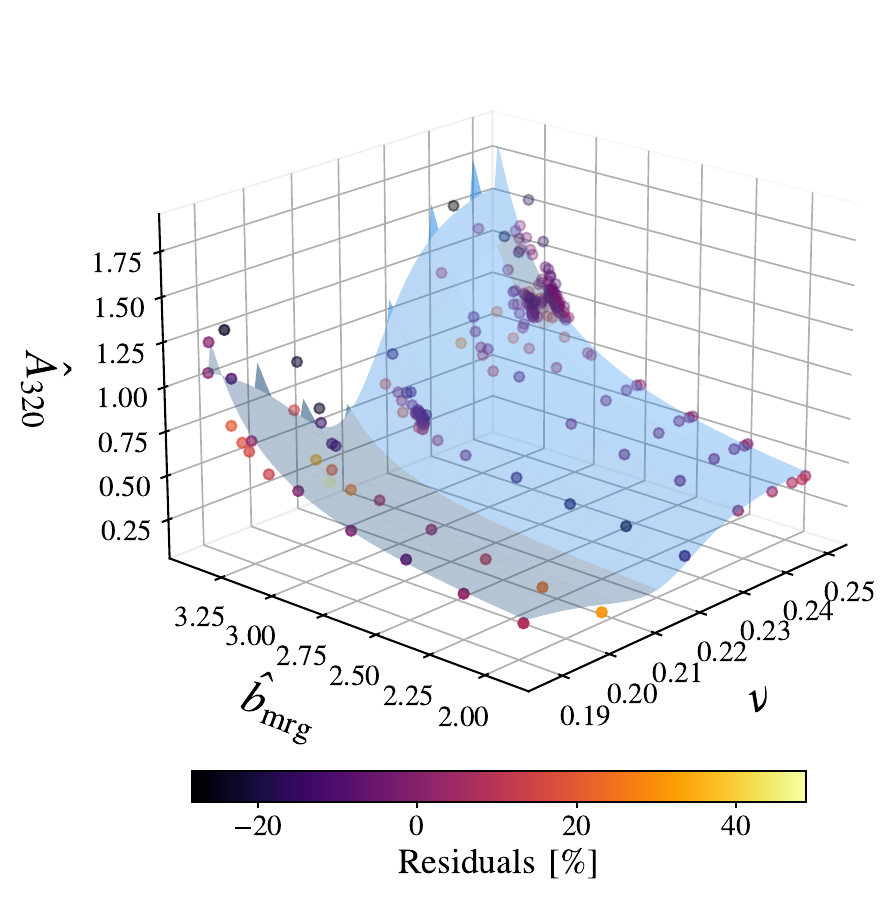}
    \includegraphics[scale=1, width=0.44\textwidth]{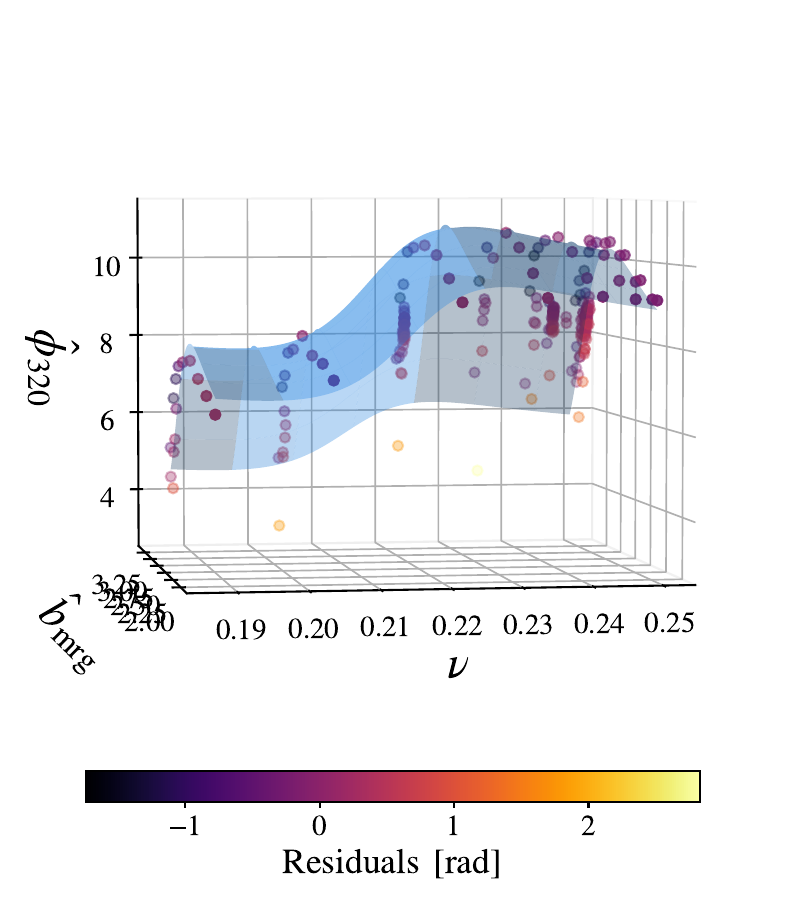}
    \includegraphics[scale=1, width=0.44\textwidth]{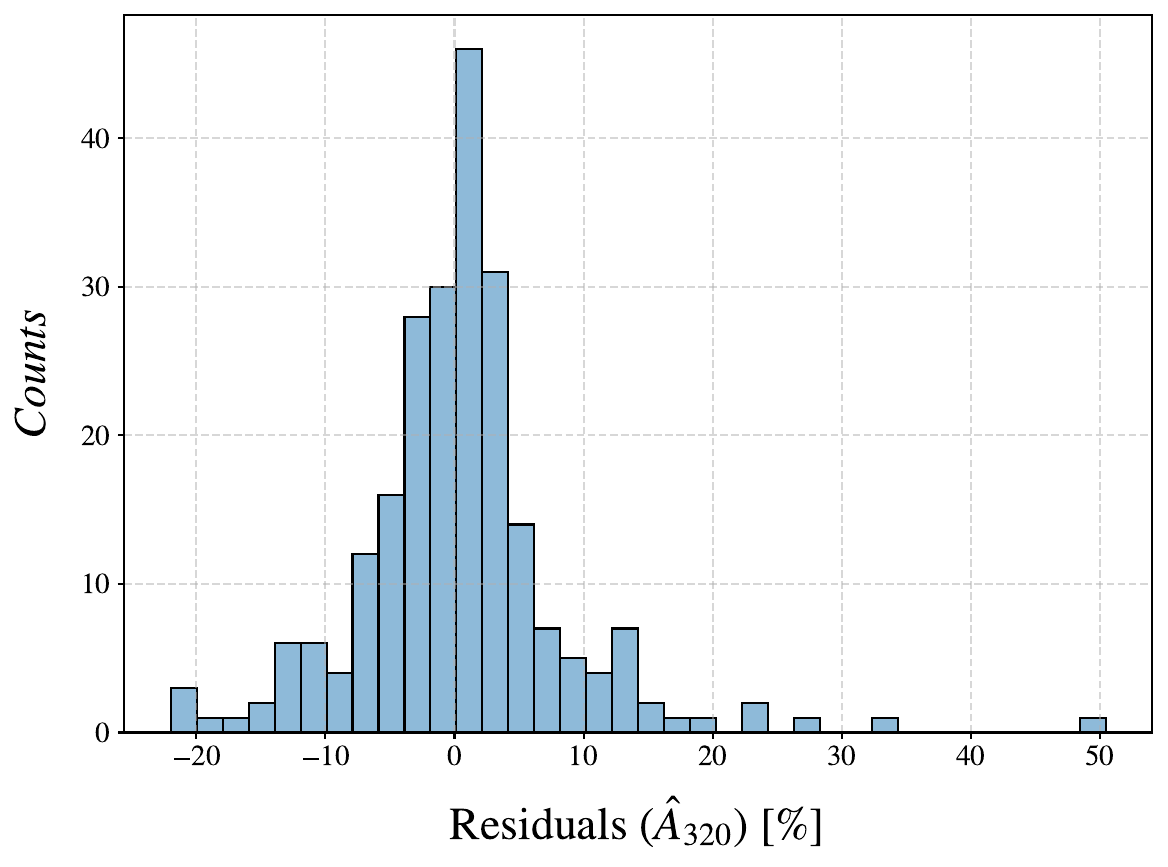}
    \includegraphics[scale=1, width=0.44\textwidth]{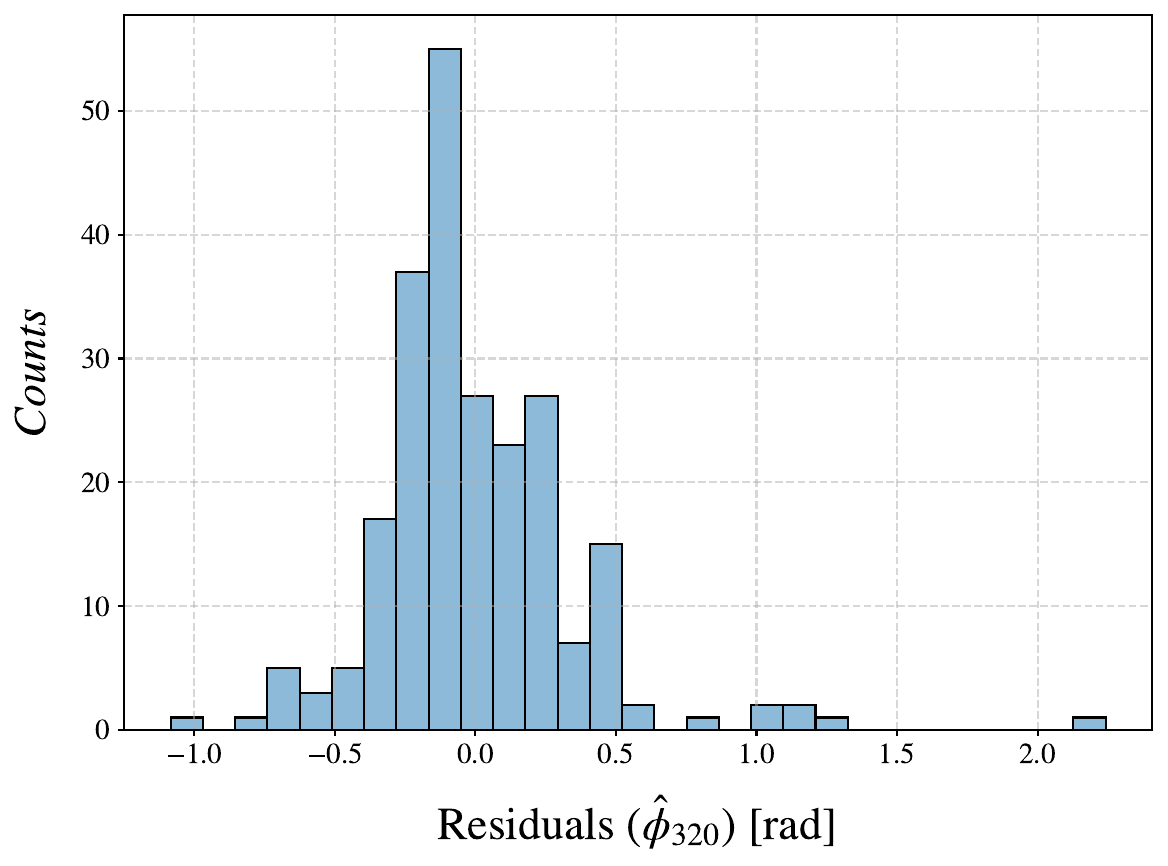}
   \caption{Parameter space (top), two-dimensional fits (middle) and three-dimensional fits (bottom) for the $(3,2,0)$ amplitude and phase parameters.
   The apparent spiky behaviour in the left central plot is a visualisation artifacts due to the surface curvature.}
   \label{fig:320_fits_unequal_mass}
\end{figure*}

The scatter and large residuals observed in the $\bmrg$ fits point to the requirement of employing the full dimensionality of the noncircular initial conditions.
Hence, Fig.~\ref{fig:residuals_3Dfit} presents the residuals of the fit constructed in terms of $(\nu, \bmrg, \Eeffmrg)$.
Agreement improves considerably, with $87\%$ of the simulations displaying an error smaller than $5\%$ on the noncircular correction to the $A_{220}$ amplitude, with $97.5\%$ smaller than $10\%$.
These numbers decrease to $85\%$ ($82\%$) below $5\%$ and $95\%$ ($93\%$) below $10\%$ for the correction to $A_{330}$ ($A_{210}$).
The $\phi_{330}$ parameter displays high accuracy, with $97\%$ of the residuals below $\rm 0.2 \, rad$, and only one simulation above $\rm 0.5 \, rad$.
The accuracy slightly decreases for $\phi_{210}$, for which only $85\%$ ($96\%$) of the simulations have residuals below $\rm 0.2 \, rad$ ($\rm 0.5 \, rad$).
Apart from a few percent outliers, such level of accuracy is comparable to the one of state-of-the-art quasi-circular models, typically around the percent-level in amplitudes and $\rm 0.2/0.3 \, rad$ in phase, see Ref.~\cite{Forteza:2022tgq}.
Fit results and residuals for the $(4,4,0)$ mode amplitudes and phases are shown in Fig.~\ref{fig:440_fits_unequal_mass}, and are similar to the ones described above, albeit less accurate as expected.
Instead, Fig.~\ref{fig:320_fits_unequal_mass} displays the results for the dominant harmonic receiving mode-mixing, $(\ell,m)=(3,2)$.
As discussed in the equal-mass results description, systematic errors in the numerical data seem to affect this mode to a larger degree, yielding a few outliers above the 20\% ($\rm 0.5 \, rad$) level for the amplitude (phase) parameters.
However, residuals for the majority of the simulations still present errors of the same order of magnitude as other modes, with $91 \%$ of the dataset carrying residuals below $10\%$ ($\rm 0.2 rad$) for the amplitude (phase) parameter.
The mass ratio dependence presents a richer structure in this case, with a non-monotonic behaviour in the amplitude and a large drop in the phase around $\nu = 0.21$.
Both these features also appear in the quasi-circular case~\cite{London:2014cma}.

In summary, fit results robustly indicate that ringdown amplitudes can increase by more than 50\% compared to the quasi-circular case for high values of the impact parameter (medium eccentricities).
A few datapoints indicate that such increase might reach even 100\%, albeit these appear as outliers compared to the bulk of the dataset. 
Additional high-quality simulations are needed to assess the robustness of such even larger increase.
The low impact parameter (highly eccentric) limit is instead smoothly represented by the fit, with a suppression of up to 50\%.

%%%%%%%%%%%%%%%%%%%%%%%%%%%%%%%%%%%%%%%%%%%%%%%%%%%%%%%%%%%%%%%%%%%%%%%%%%%%%%%%%%%%%%%%%%%%%%%%
\section{Conclusions}
\label{sec:conclusions}

Global fits for the dominant complex ringdown amplitudes associated to linear quasinormal frequencies, valid for nonspinning unequal binaries in arbitrarily eccentric orbits, were presented.
Attention was payed in considering only modes respecting time-translation symmetry, necessary when lacking a model for the transient regime, and in mitigating data-quality issues present in higher angular harmonics of available public data.
The fits were constructed in a factorised-form, based upon the highly accurate quasi-circular model of Ref.~\cite{Cheung:2023vki}, and can be straightforwardly applied to augment any exhisting quasi-circular model.

These results confirm the generic applicability and practical advantages of the parameterisation introduced in Ref.~\cite{Carullo:2023kvj} to construct models for binaries in generic orbits, previously used to obtain a closed-form representation for merger-remnant quantities.
Special focus was paid to the transition regime between high impact merger parameters (medium eccentricities) and low ones (high eccentricities), which was linked to the transition between an ``orbital-type'' and an ``infall-type'' dynamics.
The former cases display the largest increase of ringdown amplitudes compared to the quasi-circular case, up to at least $50\%$, while the latter display a significant suppression.
An interpretation of the dominant role of the impact parameter at merger was also sketched, relying on known characteristics of the plunge regime~\cite{Buonanno:2000ef,Ori:2000zn,Kuchler:2024esj}.

Future work directions include the extension to the spinning case, incorporating test-mass data to ensure a smooth extreme-mass-ratio limit, modeling the kick dependence uncovered in Ref.~\cite{Radia:2021hjs}, Bayesian global fits to mitigate the simulations' systematic uncertainties limiting the model accuracy, and parameter estimation application to spectroscopic-like analyses, together with searches of new physics.
All these extensions would undoubtedly benefit from extended explorations of the noncircular parameter space on the lines of Ref.~\cite{Albanesi:2024xus} (especially for high impact parameters and high energies), higher simulation accuracy and the availability of multiple resolutions, allowing to construct more sophisticated error models.

\acknowledgments
I am grateful to Mark Ho-Yeuk Cheung and Lionel London for constructive discussions, to Shilpa Kastha for stimulating interest in this problem and collaboration in the initial attempts to tackle it, to Jaime Redondo-Yuste, Vitor Cardoso, Rossella Gamba, Giada Caneva Santoro, Vasco Gennari, Michalis Agathos and Alex Nielsen for useful feedback on the manuscript, and to Manuela Campanelli, James Healy, Carlos Lousto and Yosef Zlochower for the creation and maintenance of the RIT public waveform database, on which this work is based.
I thank the Institut des Hautes Études Scientifiques for its warm hospitality, where part of this work was carried out.
The present research was also partly  supported by the ``\textit{2021 Balzan Prize for Gravitation: Physical and Astrophysical Aspects}'', awarded to Thibault Damour.
I also thank Emanule Berti and the Bloomberg Center for Physics and Astronomy of The Johns Hopkins University, as well as Vitor Cardoso and the Centro de Astrofisica e Gravitação of Instituto Superior Técnico for generous hospitality during the last stages of this work.
I acknowledge funding from the European Union’s Horizon 2020 research and innovation program under the Marie Sklodowska-Curie grant agreement No. 847523 ‘INTERACTIONS’, and support from the Villum Investigator program by the VILLUM Foundation (grant no. VIL37766) and the DNRF Chair program (grant no. DNRF162) by the Danish National Research Foundation.
This project has received funding from the European Union's Horizon 2020 research and innovation programme under the Marie Sklodowska-Curie grant agreement No 101131233.

\textit{Software:} contents of this manuscript have been derived using the publicly available \texttt{python} software packages: \texttt{cpnest, cython, matplotlib, numpy, pandas, pyRing, scipy, seaborn}~\cite{cpnest,cython, matplotlib,numpy,pandas,pandas_zenodo,pyRing,scipy,seaborn}.
The obtained coefficients and a \texttt{python} implementation of the fits are publicly available at: \href{https://github.com/GCArullo/noncircular_BBH_fits}{github.com/GCArullo/noncircular\_BBH\_fits}.

\clearpage

\bibliography{bibliography}

\end{document}